\documentclass[12pt,showkeys,amsmath,amssymb]{revtex4}
\usepackage{amsmath,amsfonts,amsthm,amscd,amssymb,latexsym}
\usepackage{bm}
\usepackage{dcolumn}
\usepackage{graphicx}
\usepackage{epstopdf}
\usepackage{color}
\usepackage{epsf}
\usepackage{epsfig}
\usepackage{graphicx, epic, eepic, color}
\usepackage[colorlinks,citecolor=blue,urlcolor=blue,linkcolor=blue]{hyperref}
\usepackage{float}
\usepackage[caption = false]{subfig}

\begin{document}

\title{Elliptically Polarized Plane Gravitational Waves}

\author{Masoud \surname{Molaei}$^{1}$}
\email{masoud.molaei@sharif.ir}
\author{Shant \surname{Baghram}$^{1,2}$}
\email{baghram@sharif.edu}
\author{Bahram \surname{Mashhoon}$^{1,3,4}$}
\email{mashhoonb@missouri.edu}

\affiliation{$^1$Department of Physics, Sharif University of Technology, Tehran 11155-9161, Iran\\ 
$^2$Research Center for High Energy Physics, Department of Physics, Sharif University of Technology, Tehran 11155-9161, Iran\\ 
$^3$School of Astronomy, Institute for Research in Fundamental Sciences (IPM), Tehran 19395-5531, Iran\\
$^4$Department of Physics and Astronomy, University of Missouri, Columbia,
Missouri 65211, USA\\
}

\date{\today}

\begin{abstract}
Exact plane gravitational radiation fields are presented within the framework of general relativity and their properties are described. The physics of nonlinear elliptically polarized plane gravitational waves is developed in close analogy with electromagnetic waves.  The motion of free test particles in the dynamic gravitational fields of elliptically polarized plane waves is investigated. In particular, we demonstrate the cosmic jet property of these spacetimes, namely, most timelike geodesics tend to line up in the direction of wave propagation and produce a cosmic jet whose speed asymptotically approaches the speed of light. 
\end{abstract}


\maketitle

\section{Introduction}

Gravitational radiation is a natural consequence of a field theory of the gravitational interaction. Since 2015, gravitational waves have been detected in ground-based laboratories~\cite{GW0, GW}. The results are consistent with Einstein's general relativity (GR). As in the case of electromagnetic waves, gravitational waves can carry energy, momentum and angular momentum. The LIGO-Virgo-KAGRA (LVK) collaboration has thus far reported the detection of about one hundred transient gravitational-wave signals. The Gravitational-Wave Transient Catalog, which includes the results as of January 2022, contains ninety detections of gravitational waves.  Most of the reported signals appear to be due to black-hole mergers; for the analysis of the GW signals, see, for instance,~\cite{Klimenko:2005xv, KYMM, Creswell:2017rbh, LIGOScientific:2018czr, Roulet:2019hzy, Akhshi:2021nsy, Blanchet:2023bwj, Blanchet:2023sbv, Isi:2022mbx} and the references cited therein. 
Other gravitational wave observatories are under development, such as the space-based Laser Interferometer Space Antenna (LISA). Moreover, pulsar timing residuals from an ensemble of highly stable millisecond pulsars have been used to search for a stochastic background of very low-frequency 
($\sim$ several nHz) gravitational waves~\cite{NANOGrav:2023gor}. Data from artificial satellites have also been used to place upper limits on an isotropic cosmic background of gravitational waves in the microhertz region~\cite{Mashhoon:1980pf, AnMa}.

General relativity is a nonlinear theory of gravitation  and a substantial body of literature exists on the exact solutions of GR that represent gravitational radiation; 
see~\cite{Ehlers:1962zz, Stephani:2003tm, Griffiths:2009dfa} and the references cited therein.  Further recent studies regarding exact gravitational waves are contained in~\cite{Cropp:2010zk, Harte:2015ila, Zhang:2017rno, Zhang:2017geq, Zhang:2018srn, Zhang:2019gdm, Wang:2018iig, Flanagan:2019ezo, Bini:2018gbq, Bini:2018iyu, Rosquist:2018ore, Firouzjahi:2019qmy, Dautcourt:2019fza, Hogan:2022reu, Roche:2022bcz} and the references therein. The focus of the present paper is on the physics of nonlinear elliptically polarized plane gravitational waves. To place our work in a broader context, we begin with linearized gravitational waves in GR. 

\subsection{Linear Gravitational Waves}

The spacetime metric for a linear gravitational radiation field on the background Minkowski spacetime can be written as $g_{\mu \nu} = \eta_{\mu \nu} + h_{\mu \nu}(x)$, where $x^\alpha = (t, x, y, z)$.  We use natural units throughout such that the speed of light in vacuum and Newton's constant of gravitation are set equal to unity. Furthermore, the signature of the spacetime metric is +2, greek indices run from 0 to 3, while latin indices run from 1 to 3. It is possible to impose the transverse-traceless (TT) gauge condition on  the symmetric tensor $h_{\mu \nu}$, namely,  $h^{\mu \nu}{}_{,\nu} = 0$, $h^{\mu}{}_{\mu} = 0$ and $h_{0\mu} = 0$. In this gauge, the gravitational potentials $h_{ij}(x)$ are transverse and  each satisfies the standard wave equation. The gravitational radiation field can thus be  expressed as a Fourier sum of plane monochromatic components each with frequency $\omega > 0$ and corresponding wave vector $\mathbf{k}$, $\omega =  |\mathbf{k}|$. 

Let us assume, for instance, that  the linear monochromatic plane gravitational wave of frequency $\omega$ travels in the $z$ direction.  Then, $h_{ij}$ can be written as
\begin{equation}\label{1}
(h_{ij}) = 
\begin{bmatrix}
h_{+} & h_{\times} & 0 \\
h_{\times} & - h_{+} & 0 \\
0 & 0 &0
\end{bmatrix}
\,,
\end{equation}
where 
\begin{equation}\label{2}
h_{+} = \tilde{h}_{+}\,\cos[\omega(t-z) +\varphi_{+}]\,, \qquad h_{\times} = \tilde{h}_{\times}\,\cos[\omega(t-z) +\varphi_{\times}]\,
\end{equation}
represent the plus  (``$\oplus$") and the cross  (``$\otimes$") linear polarization states of the radiation, respectively. Here, $(\tilde{h}_{+}, \varphi_{+})$ and $(\tilde{h}_{\times}, \varphi_{\times})$ are constants associated with the independent linear polarization states of the radiation field. 

Observers that remain spatially at rest in a spacetime with a metric of the form $-dt^2 + g_{ij}(x) dx^idx^j$ follow geodesic world lines. Henceforth, we consider this preferred class of geodesic observers each at rest in space with a 4-velocity vector $u^{\mu} = \delta^\mu_0$ in the gravitational radiation field under consideration here. Each fiducial observer carries an adapted orthonormal tetrad frame $e^{\mu}{}_{\hat \alpha}$,
\begin{equation}\label{3}
e^{\mu}{}_{\hat \alpha} =  \delta^\mu_\alpha -\frac{1}{2} \,h^{\mu}{}_{\alpha}(x)\,,
\end{equation}
that is parallel propagated along its world line in the TT gauge. The projection of the Riemann tensor on this tetrad frame,
\begin{equation}\label{4}
R_{\hat \alpha \hat \beta \hat \gamma \hat \delta} = R_{\mu \nu \rho \sigma}\,e^{\mu}{}_{\hat \alpha}\,e^{\nu}{}_{\hat \beta}\,e^{\rho}{}_{\hat \gamma}\,e^{\sigma}{}_{\hat \delta}\,,
\end{equation}
is the curvature of the spacetime as measured by the preferred observers.

Let us briefly digress here and mention that in an arbitrary gravitational field, we can express Eq.~\eqref{4} as a $6\times 6$ matrix with indices that range over the set $\{01,02,03,23,31,12\}$ and the result involves $3\times 3$ matrices in the form
\begin{equation}\label{5}
\left[
\begin{array}{cc}
\mathcal {E} & \mathcal{B}\cr
\mathcal{B}^{\rm T} & \mathcal{S}\cr 
\end{array}\right]\,,
\end{equation}
where $\mathcal{E}$, the gravitoelectric components of the measured curvature, and $\mathcal{S}$, the spatial components of the measured curvature, are symmetric $3\times 3$ matrices while $\mathcal{B}$, the gravitomagnetic components of the measured curvature, is traceless due to the symmetries of the Riemann tensor.   In any Ricci-flat spacetime, we find Eq.~\eqref{5} takes the form
\begin{equation}
\label{6}
\mathcal{C} = \left[
\begin{array}{cc}
\mathcal{E} & \mathcal{B}\cr
\mathcal{B} & - \mathcal{E}\cr
\end{array}
\right]\,,
\end{equation}
where $\mathcal{E}$  and $\mathcal{B}$ are $3\times 3$ symmetric and traceless matrices and are the \lq\lq electric" and \lq\lq magnetic" components of the Weyl conformal curvature tensor. 

Let us now return to the projection of the Weyl curvature tensor of the incident gravitational wave~\eqref{4}  on the tetrad frame~\eqref{3}, which to linear order in gravitational potentials coincides in this case with the natural tetrad frame $e^{\mu}{}_{\hat \alpha} = \delta^\mu_\alpha$ of the background Minkowski spacetime. Thus, the gravitoelectric components of the Weyl tensor as measured by the background inertial observers are given by 
\begin{equation}\label{7}
\mathcal{E}_{ij} = - \frac{1}{2} \frac{\partial^2 h_{ij}}{\partial t^2}\,.
\end{equation}
For Eq.~\eqref{1}, we find
\begin{equation}\label{8}
{\mathcal E}= \frac{1}{2} \,\omega^2\,\left[
\begin{array}{ccc}
 h_{+}&h_{\times} & 0\cr
h_{\times}& - h_{+} & 0\cr
0&0 & 0\cr
\end{array}
\right]\,,\qquad
{\mathcal B}= \frac{1}{2}\, \omega^2\,\left[
\begin{array}{ccc}
 h_{\times}&- h_{+} & 0\cr
- h_{+}&- h_{\times} & 0\cr
0&0 & 0\cr
\end{array}
\right]\,.
\end{equation}
Gravitational waves are detected via their tidal effects that are due to the measured components of the spacetime curvature of the radiation. It is therefore important to note here that in Eq.~\eqref{8}, the frequency of the curvature generated by the wave is the same as the wave frequency that appears in the spacetime metric.  This general correspondence in linearized GR does not hold in the nonlinear theory. 

For measurement purposes, it proves interesting to set up a quasi-inertial Fermi normal coordinate system based on the parallel-propagated tetrad frame field~\eqref{3} along the geodesic world line $\bar{x}^{\mu}(t)$ of an arbitrary fiducial  geodesic  observer fixed in space at $(\bar{x}, \bar{y}, \bar{z})$. At any given event with proper time $t$ along $\bar{x}^{\mu}(t)$, imagine the local hypersurface that is formed by spacelike geodesic curves that  are orthogonal to $\bar{x}^{\mu}(t)$. Let $x^\mu$ be the spacetime  coordinates of an event on this hypersurface that can be connected to  $\bar{x}^{\mu}(t)$ by a unique spacelike geodesic of proper length $\sigma$.  Moreover, we define $\bar{\xi}^\mu$, $\bar{\xi}^\mu u_\mu = 0$, to be the unit spacelike vector  that is tangent to the unique spacelike geodesic at event $t$.  The reference observer then assigns Fermi coordinates $X^{\hat \mu}$ to $x^\mu$ such that 
\begin{equation}\label{9}
X^{\hat 0} = t\,, \qquad X^{\hat i} = \sigma \,\bar{\xi}^\mu \,e_{\mu}{}^{\hat i}\,.
\end{equation}

Under the coordinate transformation $(t, x, y, z) \mapsto (t, X^{\hat 1}, X^{\hat 2}, X^{\hat 3})$, the spacetime metric in Fermi coordinates is given by~\cite{Chicone:2002kb}
\begin{equation}\label{10}
ds^2 = (\eta_{\hat \mu \hat \nu} + \mathfrak{h}_{\hat \mu \hat \nu})\,dX^{\hat \mu} dX^{\hat \nu}\,,
\end{equation}
where
\begin{equation}\label{11}
\mathfrak{h}_{\hat 0 \hat 0} =  - R_{\hat 0 \hat i \hat 0 \hat j}(t)\,X^{\hat i} X^{\hat j}\,,\quad \mathfrak{h}_{\hat 0 \hat i} =  - \frac{2}{3}\,R_{\hat 0 \hat j \hat i \hat k}(t)\,X^{\hat j} X^{\hat k}\,,\quad \mathfrak{h}_{\hat i \hat j} =  - \frac{1}{3}\,R_{\hat i \hat k \hat j \hat l}(t)\,X^{\hat k} X^{\hat l}\,.
\end{equation}
In expanding the Fermi metric about the Minkowski metric, we have neglected third and higher-order terms. The Fermi normal coordinate system is admissible within a cylindrical spacetime domain around the world line of the fiducial observer $\bar{x}^{\mu}(t)$. This cylinder's radius is given by an appropriate radius of curvature of spacetime~\cite{Chicone:2002kb, Chicone:2005vn}.

The physical considerations that follow involve observers that are near the fiducial observer in the Fermi frame and spatially at rest at $(X^{\hat 1},  X^{\hat 2}, X^{\hat 3})$. The structure of the Fermi metric implies that the proper time of such an observer differs from $t$, the proper time of the fiducial observer, by terms of second and higher orders in the distance away from the fiducial observer fixed at the spatial origin of Fermi coordinates.  

Let us define the gravitoelectric  potential $\hat{\Phi}_g$,
\begin{equation}\label{12}
\hat{\Phi}_g:= -\frac{1}{2} \mathfrak{h}_{\hat 0 \hat 0} = \frac{1}{2}\, R_{\hat 0 \hat i \hat 0 \hat j}\,X^{\hat i} X^{\hat j}\,, 
\end{equation}
and the gravitomagnetic vector potential $\hat{\mathbf{A}}_g$,
\begin{equation}\label{13}
(\mathbf{\hat{A}}_g)_i := -\frac{1}{2} \mathfrak{h}_{\hat 0 \hat i} =  \frac{1}{3}\,R_{\hat 0 \hat j \hat i \hat k}\,X^{\hat j} X^{\hat k}\,.
\end{equation}
Then, the gravitoelectric field $\mathbf{\hat{E}}_g$  and the gravitomagnetic field  $\mathbf{\hat{B}}_g$ can be defined in analogy with electrodynamics as
\begin{equation}\label{14}
\mathbf{\hat{E}}_g = \nabla \hat{\Phi}_g\,, \qquad  \mathbf{\hat{B}}_g = \nabla \times \mathbf{\hat{A}}_g\,.
\end{equation}
It follows that to linear order in Fermi distance, we have
\begin{equation}\label{15}
(\mathbf{\hat{E}}_g)_i = R_{\hat 0 \hat i \hat 0 \hat j}\,X^{\hat j}\,, \qquad (\mathbf{\hat{B}}_g)_i =  -\frac{1}{2}\,\epsilon_{ijk}\,R^{\hat j \hat k}{}_{\hat 0 \hat l}\,X^{\hat l}\,.
\end{equation}
Our gravitoelectromagnetic (GEM) approach here is in conformity with the treatment of the weak exterior gravitational fields of rotating astronomical bodies in GR~\cite{Bini:2021gdb}. In particular, the gravitoelectric potential reduces to the Newtonian potential in the correspondence limit. 

The fiducial observer permanently occupies the spatial origin of the Fermi coordinate system; therefore, the GEM fields vanish for the reference observer. The approximate nature of our treatment must be emphasized here. Each of the GEM quantities under consideration is expressible as a power series in the spatial distance away from the fiducial observer; however, we have only kept the first term in such a series and ignored the rest in our approximation scheme. 

Henceforth, for the sake of convenience, we assume 
\begin{equation}\label{16}
 (X^{\hat 1}, X^{\hat 2}, X^{\hat 3}) = (\widehat X,  \widehat Y,  \widehat Z)\,.
 \end{equation} 
For the incident linear gravitational wave under consideration, Eq.~\eqref{12} implies
\begin{equation}\label{17}
\hat{\Phi}_g = \frac{1}{4}\,\omega^2\,[h_{+} (\widehat{X}^2 - \widehat{Y}^2) + 2h_{\times} \widehat{X}\widehat{Y}]\,,
\end{equation}
where $h_{+} = \tilde{h}_{+}\,\cos[\omega(t-\bar{z}) +\varphi_{+}]$ and  $h_{\times} = \tilde{h}_{\times}\,\cos[\omega(t-\bar{z}) +\varphi_{\times}]$. Similarly, Eq.~\eqref{13} implies
\begin{equation}\label{18}
(\mathbf{\hat{A}}_g)_1 = \frac{1}{6}\,\omega^2\, \widehat{Z}\,(h_{+} \widehat{X} + h_{\times} \widehat{Y})\,, \quad (\mathbf{\hat{A}}_g)_2 = \frac{1}{6}\,\omega^2\, \widehat{Z}\,(h_{\times} \widehat{X} - h_{+} \widehat{Y})\,, \quad (\mathbf{\hat{A}}_g)_3 = -\frac{2}{3}\,\hat{\Phi}_g\,.
\end{equation}
The corresponding fields are then given by
\begin{equation}\label{19}
(\mathbf{\hat{E}}_g)_1 = \frac{1}{2}\,\omega^2\,(h_{+} \widehat{X} + h_{\times} \widehat{Y})\,, \qquad (\mathbf{\hat{E}}_g)_2 = \frac{1}{2}\,\omega^2\,(h_{\times} \widehat{X} - h_{+} \widehat{Y})\,,\qquad (\mathbf{\hat{E}}_g)_3 = 0\,,
\end{equation}
\begin{equation}\label{20}
(\mathbf{\hat{B}}_g)_1 = -  (\mathbf{\hat{E}}_g)_2\,, \qquad (\mathbf{\hat{B}}_g)_2 = (\mathbf{\hat{E}}_g)_1\,,\qquad (\mathbf{\hat{B}}_g)_3 = 0\,.
\end{equation}
The divergence and curl of the GEM fields $\hat{\mathbf{E}}_g$ and  $\hat{\mathbf{B}}_g$ vanish. 

In the quasi-inertial Fermi frame, an observer that is spatially at rest at $(\widehat X,  \widehat Y,  \widehat Z)$ finds that the gravitoelectric and gravitomagnetic fields are both clearly transverse to the direction of wave propagation, namely, 
the $ \widehat Z$ axis. Moreover,  $\hat{\mathbf{E}}_g \times \hat{\mathbf{B}}_g$  is in the $ \widehat Z$ direction as well and, as in electrodynamics,  the field invariants vanish, namely,   
$\hat{\mathbf{E}}^2_g - \hat{\mathbf{B}}^2_g = 0$  and $\hat{\mathbf{E}}_g \cdot \hat{\mathbf{B}}_g = 0$.  The incident wave given by Eqs.~\eqref{1} and~\eqref{2} is a superposition of the two independent linear polarization states and is thus elliptically polarized in close analogy with electrodynamics. For the special case of circular polarization, we require that in Eq.~\eqref{2},
\begin{equation}\label{21}
\tilde{h}_{+} = \tilde{h}_{\times}\,, \qquad  \varphi_{+} = \varphi_{\times} \pm \frac{\pi}{2}\,,
\end{equation}
where the upper (lower) sign refers to right (left) circularly polarized radiation.

According to the observer at $(\widehat X,  \widehat Y,  \widehat Z)$, when right (left) circularly polarized linear gravitational radiation propagates along the $\widehat Z$ direction, the gravitoelectric and gravitomagnetic fields rotate with the \emph{same} angular speed $\omega$ as the wave frequency in the positive (negative) sense about the direction of wave propagation just as in electrodynamics. This can be proved explicitly using Eqs.~\eqref{19} and~\eqref{20}. In the general case of elliptical polarization, we note that 
\begin{equation}\label{22}
\frac{d \hat{\mathbf{E}}_g}{dt}\cdot \hat{\mathbf{B}}_g = \frac{1}{4} \omega^5\, \tilde{h}_{+} \,\tilde{h}_{\times}\,\sin(\varphi_{+} - \varphi_{\times})\, (\widehat{X}^2 + \widehat{Y}^2)\,,
\end{equation}
where we use $t$ for the proper time of the observer fixed at spatial Fermi coordinates $(\widehat X,  \widehat Y,  \widehat Z)$, since we neglect higher order terms in the spatial Fermi distance in our approximation scheme. It follows that the sense of rotation of the GEM fields in the general case of elliptical polarization depends upon the sign of 
\begin{equation}\label{23}
 \tilde{h}_{+} \,\tilde{h}_{\times}\,\sin(\varphi_{+} - \varphi_{\times})\,.
\end{equation}
If this quantity is positive (negative), the fields rotate in the positive (negative) sense about the $\widehat Z$ axis and the gravitational radiation has right-handed (left-handed) polarization. This result is a generalization of Eq.~\eqref{21}, which holds for circular polarization, to the case of elliptical polarization. 

The treatment presented in this section is valid for a linear superposition of different monochromatic components all propagating in the same direction. Further investigations of the polarization properties of linear gravitational waves in GR, especially in connection with spin-rotation coupling, are contained in~\cite{BMashhoon, Ramos:2006sb, Mashhoon:2022fua, Ruggiero:2023wge}. 

We now turn to a discussion of exact elliptically polarized gravitational plane waves that are nonlinear analogues of the corresponding linear gravitational waves.

\section{Nonlinear Plane Gravitational Waves}

Consider an exact gravitational plane wave propagating along the positive $z$ direction. Introducing retarded and advanced null coordinates $u$ and $v$, respectively, by
\begin{equation}
\label{I1}
u = t-z\,, \qquad v = t+ z\,,
\end{equation}
the spacetime metric can be expressed as
\begin{equation}\label{I2}
ds^2 = - dt^2 + dz^2 + P^2(u) dx^2 + 2 R(u) dx dy +Q^2(u) dy^2\,.
\end{equation}
The coordinate system is admissible for
\begin{equation}\label{I3}
P^2 > 0\,,  \qquad \Delta := P^2(u) \,Q^2(u) - R^2(u) >0\,,
\end{equation}
where $\det(g_{\mu \nu}) = - \Delta$. The connection between $P$, $Q$ and $R$ is given by the gravitational field equations ($R_{\mu \nu} = 0$) that metric~\eqref{I2} satisfies for a pure radiation field. The Ricci-flat condition implies
\begin{equation}\label{I4}
P^2Q^2 \left(\frac{P_{,uu}}{P} + \frac{Q_{,uu}}{Q}\right) - R R_{,uu} + \frac{1}{2} R_{,u}^2 = \frac{1}{\Delta}P^2Q^2R^2\,\left(\frac{P_{,u}}{P} + \frac{Q_{,u}}{Q} - \frac{R_{,u}}{R}\right)^2\,.
\end{equation}

Gravitational fields that admit a covariantly constant null vector field $k$ with $k_{\mu ; \nu} = 0$ and $k^\mu k_\mu = 0$ describe plane-fronted gravitational waves with parallel rays ($pp$-waves) that were first considered by 
Brinkmann~\cite{Brinkmann:1925fr} and have since been extensively investigated; see~\cite{BaJe, Rosen, Bondi:1957dt, Bondi:1958aj} and the references therein. The $pp$-waves belong to the Kundt class of solutions of GR~\cite{Stephani:2003tm, Griffiths:2009dfa}. Gravitational plane waves constitute a subclass of $pp$-waves.  Therefore, spacetimes under consideration in the present work admit a covariantly constant null vector field $k$ that represents the propagation vector of the gravitational plane wave in the $z$ direction at the speed of light; moreover, as plane waves they possess five Killing 
vector fields~\cite{Bondi:1958aj}.  Indeed, the spacetime with metric~\eqref{I2} contains a null Killing vector field $k = \partial_t + \partial_z$, $k_\mu = - \partial_\mu \,u$,  that forms a nonexpanding shearfree null geodesic congruence. Two further spacelike Killing vector fields, namely,  $\xi_{1} = \partial_x$ and $\xi_{2} = \partial_y$,  are characteristic of the planar symmetry of the wave front $u = {\rm constant}$. In addition, $\xi_{1}$ and $\xi_{2}$ and $\xi_{3} = \partial_v=\tfrac{1}{2}\,k$ generate an Abelian group $G_3$ which acts in null hypersurfaces that are wave fronts given by $u= {\rm constant}$. Finally, the plane waves admit two additional Killing vector fields given by~\cite{Bondi:1958aj}
\begin{equation}
\label{I5}
\xi_{4} = x\,(\partial_t + \partial_z) + \mathcal{Q}(u)\, \partial_x + \mathcal{R}(u)\,\partial_y\,,  \qquad    \xi_{5} = y\,(\partial_t + \partial_z) + \mathcal{R}(u)\,\partial_x + \mathcal{P}(u)\,\partial_y\,,
\end{equation}
where
\begin{equation}
\label{I6}
\mathcal{P}(u)= \int^u \frac{P^2(u')}{\Delta(u')} du'\,,\qquad \mathcal{Q}(u)= \int^u \frac{Q^2(u')}{\Delta(u')} du'\,,\qquad \mathcal{R}(u)= -\int^u \frac{R(u')}{\Delta(u')} du' \,.
\end{equation}
The five Killing vectors all commute, except for $[\xi_1, \xi_4 ] = [\xi_2, \xi_5] = \partial_t + \partial_z$; moreover, they are the generators of a group $G_5$, which has as its subgroup the Abelian group $G_3$. In this connection,  see~\cite{Bondi:1958aj, Griffiths:2009dfa}, Section 24.5 of~\cite{Stephani:2003tm} and the references therein.

The main purpose of the present investigation is to elucidate the physics of elliptically polarized gravitational plane waves, which are solutions of Eq.~\eqref{I4}. These waves have been the subject of extensive investigations in Brinkmann coordinates~\cite{Brinkmann:1925fr, Stephani:2003tm, Obukhov:2003br}; however, we employ throughout the more physically transparent Baldwin-Jeffrey-Rosen coordinates~\cite{BaJe, Rosen, Griffiths:2009dfa}. In these coordinates, we develop the physics of nonlinear elliptically polarized plane waves in direct correspondence with the linearized theory.

In our general plane wave spacetime, preferred observers  are spatially at rest and follow geodesic world lines. Their natural adapted tetrads $\chi^{\mu}{}_{\hat \alpha}$ point essentially along the spacetime coordinate axes, namely,  
\begin{equation}\label{I7}
\chi_{\hat 0}=\partial_t\,,\quad  \chi_{\hat 1}=P^{-1}\partial_x\,,\quad \chi_{\hat 2}= -\frac{R}{P\,\Delta^{\frac{1}{2}}}\partial_x + \frac{P}{\Delta^{\frac{1}{2}}}\partial_y\,, \quad \chi_{\hat 3}=\partial_z\,.
\end{equation}

To determine the properties of the general plane gravitational wave, preferred observers can measure the spacetime curvature. The projection of the Riemann tensor on the natural adapted tetrad of a preferred observer is given by 
\begin{equation}\label{I8}
\mathbb{R}_{\hat \alpha \hat \beta \hat \gamma \hat \delta} = R_{\mu \nu \rho \sigma}\chi^{\mu}{}_{\hat \alpha}\chi^{\nu}{}_{\hat \beta}\chi^{\rho}{}_{\hat \gamma}\chi^{\sigma}{}_{\hat \delta}\,.
\end{equation}
For metric~\eqref{I2}, the measured curvature~\eqref{I8} together with the Ricci-flat condition~\eqref{I4}  can be expressed in the  form
\begin{equation}
\label{I9}
\mathbb{R} = \left[
\begin{array}{cc}
\mathbb{E} & \mathbb{B}\cr
\mathbb{B} & - \mathbb{E}\cr
\end{array}
\right]\,,
\end{equation}
where the gravitoelectric and gravitomagnetic components are given by
\begin{equation}
\label{I10}
\mathbb{E}= \left[
\begin{array}{ccc}
\mathbb{K}_1&\mathbb{K}_2 & 0\cr
\mathbb{K}_2&-\mathbb{K}_1 & 0\cr
0&0 & 0\cr
\end{array}
\right]\,,\qquad
\mathbb{B}= \left[
\begin{array}{ccc}
\mathbb{K}_2&-\mathbb{K}_1 & 0\cr
-\mathbb{K}_1&-\mathbb{K}_2 & 0\cr
0&0 & 0\cr
\end{array}
\right]\,.
\end{equation}
Here,
\begin{equation}
\label{I11}
\mathbb{K}_1(u)= - \frac{P_{,uu}}{P}+  \frac{R^2}{\Delta}\left(\frac{P_{,u}}{P} - \frac{1}{2} \frac{R_{,u}}{R}\right)^2\,
\end{equation}
and
\begin{align}\label{I12}
\nonumber \mathbb{K}_2(u)=&{}\frac{R}{\Delta^{\frac{1}{2}}}\left[\frac{P_{,uu}}{P} - \frac{1}{2} \frac{R_{,uu}}{R}+ \frac{1}{2} \frac{R_{,u}}{R}\left(\frac{P_{,u}}{P} + \frac{Q_{,u}}{Q}\right) - \frac{P_{,u}}{P} \, \frac{Q_{,u}}{Q}\right] \\   
&{} - \frac{R^3}{\Delta^{\frac{3}{2}}}\left(\frac{P_{,u}}{P} -\tfrac{1}{2} \frac{R_{,u}}{R}\right)\,\left(\frac{P_{,u}}{P} + \frac{Q_{,u}}{Q} - \frac{R_{,u}}{R}\right)\,.  
\end{align}

A Ricci-flat solution in GR has four algebraically independent scalar polynomial curvature invariants  given by~\cite{Stephani:2003tm}
\begin{equation}\label{I13}
\mathcal{I}_1 = R_{\mu \nu \rho \sigma}\,R^{\mu \nu \rho \sigma} - i R_{\mu \nu \rho \sigma}\,R^{*\,\mu \nu \rho \sigma}\,
\end{equation}  
and
\begin{equation}\label{I14}
\mathcal{I}_2 = R_{\mu \nu \rho \sigma}\,R^{\rho \sigma \alpha \beta}\,R_{\alpha \beta}{}^{\mu \nu} + i R_{\mu \nu \rho \sigma}\,R^{\rho \sigma \alpha \beta}\,R^{*}{}_{\alpha \beta}{}^{\mu \nu}\,.
\end{equation}
Here, $R^*_{\mu \nu \rho \sigma} = \tfrac{1}{2}\,e_{\mu \nu \alpha \beta}\,R^{\alpha \beta}{}_{\rho \sigma}$ is the dual curvature tensor and $e_{\mu \nu \rho \sigma} = (-g)^{\frac{1}{2}}\,\epsilon_{\mu \nu \rho \sigma}$ is the Levi-Civita tensor with alternating symbol $\epsilon_{\mu \nu \rho \sigma}$ such that $\epsilon_{0123} := 1$. We find that $\mathcal{I}_1$ and $\mathcal{I}_2$ both vanish, since gravitational plane waves are of type N in the Petrov classification; that is, the four principal directions of the Weyl tensor coincide and are parallel to the direction of propagation of the plane wave.  It is interesting to note that in a general spacetime, it is possible to express $\mathcal{I}_1$ and $\mathcal{I}_2$ in terms of the measured components of the curvature tensor, namely, $\mathcal{E}$, $\mathcal{B}$ and $\mathcal{S}$ defined in Eq.~\eqref{5}. For instance, 
\begin{equation}\label{I15}
\frac{1}{4}\,R_{\mu \nu \rho \sigma}\,R^{\mu \nu \rho \sigma} = {\rm tr}(\mathcal {E}^2 - 2 \mathcal {B}^T\,\mathcal {B}+ \mathcal {S}^2)\,,
\end{equation}
\begin{equation}\label{I16}
\frac{1}{16}\,R_{\alpha \beta}{}^{\mu \nu}\,R^{\alpha \beta \rho \sigma}\,e_{\mu \nu \rho \sigma} = {\rm tr}(\mathcal {E}\,\mathcal {B} - \mathcal {B}\, \mathcal {S})\,.
\end{equation}
In any Ricci-flat spacetime, ${\mathcal E} = - {\mathcal S}$  and ${\mathcal B}$ are $3\times 3$ symmetric and traceless matrices and are the ``electric" and ``magnetic" components of the Weyl tensor. In this case, the curvature invariants~\eqref{I15} and~\eqref{I16} divided by two take the forms ${\rm tr}(\mathcal {E}^2 - \mathcal {B}^2)$ and ${\rm tr}(\mathcal {E}\,\mathcal {B})$, respectively, that are familiar from electrodynamics. These vanish for null fields in electrodynamics and they vanish here for plane gravitational waves via Eq.~\eqref{I10}.  

If $R = 0$, the tetrad frame field~\eqref{I7} is parallel propagated; otherwise,  it is necessary to find the acceleration tensor $\Phi_{\hat \alpha \hat \beta}$ for the class of preferred observers; that is, 
\begin{equation}\label{I17}
\frac{D \chi^{\mu}{}_{\hat \alpha}}{dt} = \Phi_{\hat \alpha}{}^{\hat \beta} \, \chi^{\mu}{}_{\hat \beta}\,,
\end{equation}
where $t$ is the fiducial observer's proper time. The acceleration tensor should only contain a rotation that would be proportional to $R$. In fact, it is straightforward to show that the only nonzero components of the acceleration tensor are given by
\begin{equation}\label{I18}
\Phi_{\hat 1 \hat 2} = - \Phi_{\hat 2 \hat 1} = \Omega(u)\,, \qquad \Omega (u) = \frac{1}{2}\Delta^{-\frac{1}{2}}\,P^{2}\,\left(\frac{R}{P^2}\right)_{,u}\,.
\end{equation}
Consequently, the spatial frame in Eq.~\eqref{I7} rotates relative to the parallel transported frame along the geodesic world line with angular velocity $\boldsymbol{\Omega}=\Omega(u)\,\hat{z}$.

The preferred observers are at fixed $(x, y, z)$ locations in space; therefore, let 
\begin{equation}\label{I19}
\Theta = \int^u \Omega (u') \,du' \,;  
\end{equation}
then, the tetrad system $e^{\mu}{}_{\hat \alpha}$ that is parallel propagated along the geodesic path of an arbitrary preferred observer is given by $e_{\hat 0} = \chi_{\hat 0}=\partial_t$, $e_{\hat 3} = \chi_{\hat 3}=\partial_z$ and 
\begin{equation}\label{I20}
e_{\hat 1} = \chi_{\hat 1}\,\cos \Theta  - \chi_{\hat 2}\,\sin \Theta\,, \qquad  e_{\hat 2} =  \chi_{\hat 1}\,\sin \Theta + \chi_{\hat 2}\,\cos \Theta\,.  
\end{equation}
Under this rotation, the components of the curvature tensor~\eqref{4} measured by the parallel-propagated tetrad frame~\eqref{I20} have the same form as in Eqs.~\eqref{I9}--\eqref{I10}, except that
 $(\mathbb{E}, \mathbb{B}) \mapsto (\mathcal{E}, \mathcal{B})$ and $(\mathbb{K}_1, \mathbb{K}_2) \mapsto (\mathcal{K}_1, \mathcal{K}_2)$, where
\begin{equation}\label{I21}
\mathcal{K}_1 = \mathbb{K}_1 \cos 2\Theta - \mathbb{K}_2 \sin 2\Theta\,, \qquad   \mathcal{K}_2 = \mathbb{K}_1 \sin 2\Theta + \mathbb{K}_2 \cos 2\Theta\,.
\end{equation}
The curvature components $(\mathcal{K}_1, \mathcal{K}_2)$ are functions of the proper time of the fiducial observer. 

As in the case of linearized plane gravitational waves, we need to establish a quasi-inertial Fermi coordinate system on the basis of the parallel-transported tetrad frame $e^{\mu}{}_{\hat \alpha}$ along the geodesic world line of a fiducial observer that is spatially at rest in the spacetime under consideration. The fiducial observer is then permanently fixed at the spatial origin of the Fermi coordinate system. The procedure that we follow is the same as in Eqs.~\eqref{9}--\eqref{11}, except that the curvature components now belong to metric~\eqref{I2}. The gravitational potentials in the Fermi system are expressible as power series in the spatial distance; however, in our approximation scheme, only the first term in the series is taken into consideration. Thus, the gravitoelectric potential~\eqref{12} of the general plane wave is given by 
\begin{equation}\label{I22}
\hat{\Phi}_g = \frac{1}{2}\,\mathcal{K}_1 (\widehat{X}^2 - \widehat{Y}^2) + \mathcal{K}_2 \widehat{X}\widehat{Y}\,,
\end{equation}
while the gravitomagnetic vector potential~\eqref{13} can be expressed as
\begin{equation}\label{I23}
\hat{\mathbf{A}}_g = \frac{1}{3}\,\left[(\mathcal{K}_1\widehat{X} + \mathcal{K}_2 \widehat{Y})\widehat{Z}, ~(\mathcal{K}_2\widehat{X} - \mathcal{K}_1 \widehat{Y})\widehat{Z}, ~-2\,\hat{\Phi}_g\right]\,.
\end{equation}
Defining GEM fields as in Eqs.~\eqref{14} and~\eqref{15}, we find
\begin{equation}\label{I24}
\hat{\mathbf{E}}_g = \left(\mathcal{K}_1\widehat{X} + \mathcal{K}_2 \widehat{Y}, ~\mathcal{K}_2\widehat{X} - \mathcal{K}_1 \widehat{Y}, ~0\right)\,,
\end{equation}
\begin{equation}\label{I25}
\hat{\mathbf{B}}_g = \left(-\mathcal{K}_2\widehat{X} + \mathcal{K}_1 \widehat{Y}, ~\mathcal{K}_1\widehat{X} + \mathcal{K}_2 \widehat{Y}, ~0\right)\,.
\end{equation}

Let us note that by replacing $\mathcal{K}_1$ with $\tfrac{1}{2}\,\omega^2\,h_{+}$ and $\mathcal{K}_2$ with $\tfrac{1}{2}\,\omega^2\,h_{\times}$, we recover the corresponding quantities for linearized waves. The GEM fields are clearly transverse to the direction of wave propagation $\widehat Z$. Furthermore,  the general GEM fields according to the observer fixed at $(\widehat X,  \widehat Y,  \widehat Z)$  are null in the sense that  $\hat{\mathbf{E}}^2_g - \hat{\mathbf{B}}^2_g = 0$  and $\hat{\mathbf{E}}_g \cdot \hat{\mathbf{B}}_g = 0$, as can be easily checked using Eqs.~\eqref{I24} and~\eqref{I25}. In addition, $\hat{\mathbf{E}}_g \times \hat{\mathbf{B}}_g$  is in the $\widehat Z$ direction, which is the direction of wave propagation. The GEM fields  have vanishing divergence and curl in our approximation scheme, where the spacetime metric in the Fermi frame is equal to the Minkowski metric up to terms of second order and higher in the spatial Fermi coordinates.  The GEM fields vanish for the fiducial observer as a consequence of Einstein's principle of equivalence. However, away from the origin of spatial Fermi coordinates but close to it, the GEM fields grow linearly with distance and the results given here are valid to lowest order in the spatial Fermi coordinates.  In any case, the Fermi coordinates are valid only within the domain of admissibility characterized by the radius of curvature of spacetime~\cite{Chicone:2002kb, Chicone:2005vn}. 

Nonlinear waves in GR carry energy, momentum and angular momentum. We can deal with these concepts within the Fermi frame via a certain averaging procedure that has been described in detail in~\cite{Mashhoon:1996wa, Mashhoon:1998tt, Mashhoon:2000he, Mashhoon:2003ax, Bini:2018lhh} and has been applied to the study of rotating gravitational waves~\cite{Bahram, Mashhoon:2000zi}. The result is the GEM energy-momentum tensor (i.e., the super-energy-momentum tensor constructed via the Bel-Robinson tensor). A brief description of this approach is contained in Appendix A. The GEM energy-momentum tensor
$\mathcal{T}_{\hat \alpha \hat \beta}$ is defined with respect to geodesic fiducial observers that carry the nonrotating tetrad system $e^{\mu}{}_{\hat \alpha}$ along their world lines and is given by
\begin{equation}\label{I26}
\mathcal{T}_{\hat \alpha \hat \beta} = \ell^2\, T_{\hat \alpha \hat \beta \hat 0 \hat 0}\,,
\end{equation}
where $\ell$ is a constant intrinsic lengthscale associated with the gravitational field and  $T_{\mu \nu \rho \sigma}$ is the Bel-Robinson tensor; see Appendix A. The appearance of $\ell$ is a natural consequence of the averaging procedure that is involved in the derivation of  Eq.~\eqref{I26} in the Fermi frame; that is, within the GEM framework, the gravitational energy-momentum tensor is defined in analogy with electrodynamics up to a positive constant multiplicative factor proportional to $\ell^2$.   It follows from the results presented in Appendix A that we can write
\begin{equation}\label{I27}
\mathcal{T}^{\mu \nu} = \rho_g \, k^\mu\, k^\nu\,,\qquad \rho_g   = 2\ell^2\,(\mathcal{K}_1^2 + \mathcal{K}_2^2)\,,
\end{equation}
where $k = \partial_t + \partial_z$. That is, the GEM energy-momentum tensor of a  general elliptically polarized plane gravitational wave behaves like the energy-momentum tensor of a bundle of null rays with energy density given by $\rho_g$. Let us note that rotation~\eqref{I21} that maps $(\mathbb{K}_1, \mathbb{K}_2)$ to   $(\mathcal{K}_1, \mathcal{K}_2)$ implies
\begin{equation}\label{I28}
 \rho_g   = 2\ell^2\,(\mathbb{K}_1^2 + \mathbb{K}_2^2)\,,
\end{equation}
which can simplify the calculation of the radiation energy density. For linear waves of the previous section, for instance, we have $\rho_g = \tfrac{1}{2} \ell^2 \omega^4 (h^2_{+} + h^2_{\times})$, which agrees with the result obtained from the Landau-Lifshitz energy-momentum pseudotensor~\cite{L+L, Mashhoon:1980pf}  for $\ell = (\sqrt{8\pi}\, \omega)^{-\frac{1}{2}}$.

Finally, relative to the observer in the Fermi frame fixed at $(\widehat X,  \widehat Y,  \widehat Z)$, the GEM fields rotate as the wave propagates along the $\widehat Z$ direction. In our approximation scheme,  the actual proper time of the observer that is spatially at rest at $(\widehat X,  \widehat Y,  \widehat Z)$ differs from $t$ by terms of second and higher orders in the spatial Fermi coordinates; therefore, to lowest order, we can approximate this proper time by $t$.   To determine the sense of rotation of the GEM fields, let us then consider 
\begin{equation}\label{I29}
\frac{d \hat{\mathbf{E}}_g}{dt}\cdot \hat{\mathbf{B}}_g = \left(\mathcal{K}_1\, \frac{d \mathcal{K}_2}{dt} -  \frac{d \mathcal{K}_1}{dt}\,\mathcal{K}_2\,\right) (\widehat{X}^2 + \widehat{Y}^2)\,,
\end{equation}
which is a generalization of Eq.~\eqref{22}. From  
 \begin{equation}\label{I30}
 \mathcal{K}_1\, \frac{d \mathcal{K}_2}{dt} -  \frac{d \mathcal{K}_1}{dt}\,\mathcal{K}_2 =  \mathcal{K}_1^2\, \frac{d}{dt}\left(\frac{\mathcal{K}_2}{\mathcal{K}_1}\right)\,,
\end{equation}
we find that when the rate of change of $\mathcal{K}_2/ \mathcal{K}_1$ is positive (negative), the fields rotate in the positive (negative) sense about the $\widehat Z$ axis  and the gravitational radiation field exhibits right-handed (left-handed) polarization. 

\subsection{Geodesic Motion in Fermi Coordinates}

To study the influence of gravitational radiation on the motion of free test particles in an invariant setting, we need to solve the geodesic equation in the Fermi coordinate system. Let $\mathcal{U}^{\hat \mu}$ be the 4-velocity vector of a free test particle; then,  
\begin{equation}\label{G1}
\mathcal{U}^{\hat \mu} := \frac{dX^{\hat \mu}}{d\vartheta} = \widehat{\Gamma}\,(1, \widehat{\mathbf{V}})\,,
\end{equation}
where $\vartheta$ is its proper time and  the Lorentz factor  $\widehat{\Gamma}$ is given by $g_{\hat \mu \hat \nu}\,\mathcal{U}^{\hat \mu}\mathcal{U}^{\hat \nu} = -1$, namely, 
\begin{equation}\label{G2}
\widehat{\Gamma}^{-2} = - g_{\hat 0 \hat 0}-2\, g_{\hat 0 \hat i}\,V^{\hat i} - g_{\hat i \hat j}\,V^{\hat i}\,V^{\hat j}\,. 
\end{equation}
The timelike geodesic equation 
\begin{equation}\label{G3}
\frac{d^2 X^{\hat \mu}}{d\vartheta^2}+\Gamma^{\hat \mu}{}_{\hat \alpha \hat \beta}\,\frac{dX^{\hat \alpha}}{d\vartheta}\,\frac{dX^{\hat \beta}}{d\vartheta}=0\,
\end{equation}
can be separated into its temporal and spatial components and after some algebra one obtains the reduced geodesic equation~\cite{Chicone:2002kb}
\begin{equation}\label{G4}
\frac{d^2 X^{\hat i}}{dt^2}+\left(\Gamma^{\hat i}{}_{\hat \alpha \hat \beta}-\Gamma^{\hat 0}{}_{\hat \alpha \hat \beta}V^{\hat i} \right) \frac{dX^{\hat \alpha}}{dt}\frac{dX^{\hat \beta}}{dt}=0\,,
\end{equation}
where $t$ is the temporal coordinate in the Fermi frame and is the proper time of the fiducial observer in our treatment. The spacetime metric at the location of the fiducial observer is Minkowskian; therefore, Fermi velocity $\widehat{\mathbf{V}}$ of the test particle at the origin of spatial Fermi coordinates  must be such that $|\widehat{\mathbf{V}}| \le 1$ at $\widehat{\mathbf{X}}=0$. The reduced geodesic equation~\eqref{G4} holds for a null geodesic as well, provided  $\widehat{\Gamma}$ diverges, namely,
\begin{equation}\label{G5}
 g_{\hat 0 \hat 0} +2\, g_{\hat 0 \hat i}\,V^{\hat i} + g_{\hat i \hat j}\,V^{\hat i}\,V^{\hat j} = 0\,. 
\end{equation}
The reduced geodesic equation~\eqref{G4} constitutes the generalized Jacobi equation~\cite{Chicone:2002kb}. 

To write down explicitly the components of the reduced geodesic equation in the present case, we drop the hats and let
\begin{equation}\label{G6}
 \widehat{\mathbf{X}} \to (X, Y, Z)\,, \qquad      \widehat{\mathbf{V}} \to (V_x, V_y, V_z)\,
\end{equation}
in order to simplify matters. The Fermi connection coefficients for metric~\eqref{I2} are given in Appendix B. Moreover, we need to define
\begin{equation}\label{G7}
 \mathcal{V}_1 := \mathcal{K}_1 V_x + \mathcal{K}_2 V_y\,, \qquad  \mathcal{V}_2 := \mathcal{K}_2 V_x - \mathcal{K}_1 V_y\, 
\end{equation}
and
\begin{equation}\label{G8}
\mathcal{W} := (X  \mathcal{V}_1 + Y  \mathcal{V}_2)(1-\tfrac{1}{3} V_z) + \tfrac{1}{3} Z (V_x \mathcal{V}_1 + V_y \mathcal{V}_2)\,. 
\end{equation}
Then, the equations of motion are given by
\begin{equation}\label{G9}
 \frac{dV_x}{dt} + (\mathcal{K}_1 X + \mathcal{K}_2 Y)(1-2V_z + \tfrac{2}{3}V_z^2) -  \tfrac{2}{3} Z V_z \mathcal{V}_1 - 2 V_x \mathcal{W} = 0\,, 
\end{equation}
\begin{equation}\label{G10}
 \frac{dV_y}{dt} + (\mathcal{K}_2 X - \mathcal{K}_1 Y)(1-2V_z + \tfrac{2}{3}V_z^2) -  \tfrac{2}{3} Z V_z \mathcal{V}_2 - 2 V_y \mathcal{W} = 0\,, 
\end{equation}
\begin{equation}\label{G11}
 \frac{dV_z}{dt} + 2 (1- V_z) \mathcal{W} = 0\,.
\end{equation}
The system of Eqs.~\eqref{G9}--\eqref{G11} is invariant under the transformations $(X, Y, Z) \mapsto (Y, X, Z)$ and $(\mathcal{K}_1, \mathcal{K}_2) \mapsto (-\mathcal{K}_1, \mathcal{K}_2)$; furthermore, these equations generalize to the case of elliptical polarization previous work that involved nonlinear plane waves with plus ($\oplus$) polarization~\cite{Chicone:2002kb}.  The fiducial geodesic observer at the origin of Fermi coordinates provides a trivial solution of these equations for $(X, Y, Z) = (0, 0, 0)$, while $(X, Y, Z) = (0, 0, t + {\rm constant})$ is the solution representing the null geodesic moving in the direction of the radiation.   In our approximation scheme, Eq.~\eqref{G11} has the exact solution $V_z = 1$ and $Z = t + {\rm constant}$.   

Let us note that for  purely  transverse motion, namely, for motion purely in the $X$ direction and separately in the $Y$ direction,  Eqs.~\eqref{G9} and~\eqref{G10} reduce to
\begin{equation}\label{G12}
 \frac{dV_x}{dt} + \mathcal{K}_1\,(1-2V^2_x) X = 0\,, \qquad  \frac{dV_y}{dt} - \mathcal{K}_1\,(1-2V^2_y) Y = 0\,, 
\end{equation}
respectively. These equations have rest points at $(X, V_x) = (0, 0)$ and $(Y, V_y) = (0, 0)$, respectively; that is, there is no motion if the initial speed is zero at the origin of Fermi coordinates. Moreover, they each have an exact solution involving uniform rectilinear motion at the constant critical speed of $V_c = 1/\sqrt{2}$, which is  $\approx 0.7$ of the speed of light. Assuming that $\mathcal{K}_1$ behaves sinusoidally in time $t$, numerical evidence for  Hamiltonian chaos in Eq.~\eqref{G12}  has been presented in~\cite{Chicone:2002kb}. On the other hand, for motion purely along  the  direction of wave propagation, we find $dV_z/dt = 0$, so that we have uniform motion along the $Z$ axis at the level of approximation for the Fermi frame under consideration here. 

Keeping the velocity terms in Eqs.~\eqref{G9}--\eqref{G11} only to first order, the geodesic equations of motion in the Fermi frame can be written in a form that is reminiscent of the Lorentz force law~\cite{Mashhoon:2000he, Mashhoon:2003ax}; that is, 
\begin{equation}\label{G13}
 \frac{d^2\widehat{\mathbf{X}}}{dt^2} +  \hat{\mathbf{E}}_g + 2 \,\widehat{\mathbf{V}} \times \hat{\mathbf{B}}_g = 0\,, 
\end{equation}
where the gravitoelectric and gravitomagnetic fields are given by Eqs.~\eqref{I24} and~\eqref{I25}.

Further properties of the equations of motion in Fermi coordinates will be discussed in Section IV.  

\subsection{Polarization}

The nature of the transverse polarization of the wave depends on the spatial frame of the observer. To illustrate this point, consider a simple rotation in the $(x, y)$ plane by a constant angle  $\check{\theta}$, namely, 
\begin{equation}\label{H1}
\check{x} = x\,\cos \check{\theta} + y\, \sin \check{\theta}\,, \qquad \check{y} = - x\, \sin \check{\theta} + y\,\cos \check{\theta}\,.
\end{equation}
Under this rotation, $(P, Q, R) \mapsto (\check{P}, \check{Q}, \check{R})$ such that $\check{P}^2 + \check{Q}^2 = P^2 + Q^2$ and 
\begin{equation}\label{H2}
\frac{1}{2}(\check{P}^2 - \check{Q}^2) = \frac{1}{2} (P^2-Q^2)\,\cos 2 \check{\theta} + R\, \sin 2\check{\theta}\,, \qquad \check{R} = - \frac{1}{2} (P^2-Q^2)\,\sin 2 \check{\theta} + R\,\cos 2\check{\theta}\,.
\end{equation}
Moreover, this transformation is such that $\check{\Delta} = \check{P}^2\, \check{Q}^2 - \check{R}^2 = \Delta$. For the linearized gravitational radiation of Section I, we find
\begin{equation}\label{H3}
\check{h}_{+}= h_{+}\,\cos 2 \check{\theta} + h_{\times}\, \sin 2\check{\theta}\,, \qquad \check{h}_{\times} = - h_{+}\,\sin 2 \check{\theta} + h_{\times}\,\cos 2\check{\theta}\,.
\end{equation}
One can check that $(h_{+}, h_{\times}) \mapsto (\check{h}_{+}, \check{h}_{\times}) = (h_{\times}, -h_{+})$ for $\check{\theta} = \tfrac{1}{4}\pi$; that is, the two independent linear polarization states are related to each other by a $45^\circ$ rotation. 

It is important to recognize that the polarization quantities $(P, Q, R)$ refer to fiducial observers with adapted tetrads~\eqref{I7} whose spatial axes essentially correspond to the  spatial coordinate axes, while  $(\check{P}, \check{Q}, \check{R})$ correspond to observers that are spatially at rest but have adapted tetrads that can be obtained from~\eqref{I7} by a constant rotation of angle $\check{\theta}$. For nonlinear gravitational plane waves, the type of polarization can be determined based on the analogy with linear gravitational waves. 


\section{Elliptical Polarization}

In this section, we search for exact elliptically polarized plane gravitational waves that are solutions of the gravitational field Eq.~\eqref{I4}, which is an ordinary differential equation involving three functions $(P, Q, R)$ of $u$. To solve Eq.~\eqref{I4}, we need to reduce judiciously the three functions to one. To this end, we proceed by trial and error.  Suppose, for instance, that $P(u)$ is a linear function of $u$ and $Q/P$ as well as $R/P^2$ are constants; then,  Eq.~\eqref{I4} is satisfied, but the solution represents flat spacetime as $\mathbb{K}_1$ and $\mathbb{K}_2$ both vanish in this case. We therefore turn to curved spacetimes that represent elliptically polarized plane waves. 

\subsection{Solutions with $P =  \Psi (R)$ and $PQ = 1$}

Let us first assume that $P(u)$ and $Q(u)$ are known functions of $R(u)$. Specifically, 
\begin{equation}\label{U1}
P =  \Psi (R)\,, \qquad    Q = \frac{1}{\Psi (R)}\,.
\end{equation}
In this case, 
\begin{equation}\label{U2}
\frac{P_{,u}}{P} + \frac{Q_{,u}}{Q} = 0\,, \qquad \frac{P_{,uu}}{P} + \frac{Q_{,uu}}{Q} = \frac{2}{\Psi^2}\left(\frac{d\Psi }{dR}\right)^2 \,R^2_{,u}\,.
\end{equation}
Therefore, Eq.~\eqref{I4} can be written as
\begin{equation}\label{U3}
(2f^2 + \tfrac{1}{2}) R_{,u}^2 - R\,R_{,uu} = \frac{R_{,u}^2}{1-R^2}\,, \qquad  f(R) := \frac{1}{\Psi}\left(\frac{d\Psi }{dR}\right)\,.
\end{equation}
The coordinate system is admissible for $\Delta > 0$; hence,  $R^2 < 1$ and a trivial solution is obtained for constant $R$. Henceforth, we assume that $R$ is not a constant; writing $R_{,uu}  =  R_{,u}\, dR_{,u}/dR$, Eq.~\eqref{U3} can be expressed as a differential equation for $R_{,u}$ in terms of $R$ and can be reduced to quadratures. 

For $f(R) = 0$, we recover a solution for the case of cross ($\otimes$) polarization. Next, we let $f(R) = \mathfrak{a}$, where $\mathfrak{a} \ne 0$ is a constant; in this case,  the general solution is
\begin{equation}\label{U4}
    \left(\frac{2 \,R^{\frac{3}{2}-2 \mathfrak{a}^2}} {3-4 \mathfrak{a}^2}\right) \, _2F_1(\frac{1}{2},\frac{3}{4}-\mathfrak{a}^2;\frac{7}{4}-\mathfrak{a}^2; R^2)=\eta u+ \varphi\,,
\end{equation}
where  ${_2}F_{1}$ is the hypergeometric function~\cite{A+S}, while $\eta$ and $\varphi$ are integration constants. 
An interesting special case is obtained for $\mathfrak{a} = \pm \tfrac{1}{2}$, namely, 
\begin{equation}\label{U5}
R = \sin(\eta u + \varphi)\,, \quad P =  e^{\pm \tfrac{1}{2}\,R}\,, \quad Q = e^{\mp \tfrac{1}{2}\,R}\,.
\end{equation}
For $\varphi \mapsto \varphi + \pi$ and $y \mapsto -y$, the upper sign in Eq.~\eqref{U5} changes to the lower sign; therefore, we henceforth employ the upper sign without any loss in generality.  For this special solution, the measured components of curvature can be obtained from $\mathbb{K}_1$ and $\mathbb{K}_2$ in Eqs.~\eqref{I11} and~\eqref{I12} and the results are
\begin{equation}\label{U6}
\mathbb{K}_1=\frac{\eta^2}{2}\sin^2(\eta u+ \varphi)\,,\qquad \mathbb{K}_2=\frac{\eta^2}{4}\sin(2\,\eta u+ 2\, \varphi)\,.
\end{equation}
Therefore, there is no curvature singularity here. Moreover, the energy density of the wave is given by $\rho_g = \tfrac{1}{2} \ell^2 \eta^4\, R^2$.  The coordinate system becomes inadmissible for $R = \sin(\eta u + \varphi) = 1$.   Finally, we mention that in this special case the nonzero components of the acceleration tensor~\eqref{I18} can be obtained from
\begin{equation}\label{U7}
\Omega = \tfrac{1}{2} \eta \left[1-\sin(\eta u + \varphi)\right]\,
\end{equation}
and the corresponding angle of rotation~\eqref{I19} is given by
\begin{equation}\label{U8}
\Theta = \tfrac{1}{2} [\eta u + \varphi + \cos(\eta u + \varphi) + \varphi']\,,
\end{equation}
where $\varphi'$ is an integration constant that depends on the particular fiducial observer under consideration here. We return to this  special nonsingular solution in Section IV.

\subsection{Solutions with $P = \phi\,\psi$, $Q = \phi\,\psi^{-1}$ and  $R = \phi^2 \,r(u)$}

Instead of $PQ = 1$, as in the previous subsection, we now assume $PQ = \phi^2(u) >0$. Indeed. 
\begin{equation}\label{V1}
    P= \phi(u)\, \psi(u)\,, \qquad  Q= \phi(u)\, \psi^{-1}(u)\,, \qquad R= \phi^2(u)\,r(u)\,,
\end{equation}
where $\psi(u) > 0$ and  $\Delta = (1-r^2)\phi^4(u) > 0$. We find
\begin{equation}\label{V2}
 \frac{P_{,u}}{P} +  \frac{Q_{,u}}{Q} -  \frac{R_{,u}}{R} = -  \frac{r_{,u}}{r}\,,
\end{equation}
\begin{equation}\label{V3}
  \frac{P_{,uu}}{P}+\frac{Q_{,uu}}{Q}=2 \frac{\phi_{,uu}}{\phi}+2 \left(\frac{\psi_{,u}}{\psi}\right)^{2}\,, 
\end{equation}
\begin{equation}\label{V4}
 R_{,u} = \phi^2 \,r_{,u} + 2 \phi\, r\, \phi_{,u}\,, \qquad R_{,uu}= \phi^2\, r_{,uu} + 4\phi \phi_{,u} r_{,u}+2 r \phi_{,u}^2 +2 \phi\,r\,\phi_{,uu}\,.
\end{equation}
Using these relations, Eq.~\eqref{I4} takes the form
\begin{equation}\label{V5}
    (1-r^2)\frac{\phi_{,uu}}{\phi}+\left(\frac{\psi_{,u}}{\psi}\right)^2 - \frac{1}{2}\,r\,r_{,uu} - r\,r_{,u}\, \frac{\phi_{,u}}{\phi} -\frac{1}{4}\left(\frac{1+r^2}{1-r^2}\right)\,r^2_{u}=0\,.
\end{equation}

Let us first assume that $r_{,u} \ne 0$. Then, Eq.~\eqref{V5} can be reduced to quadratures if we express it as a differential equation for $r_{,u}$ in terms of $r$ and at the same time suppose that $d (\ln \phi)/dr$ and $d( \ln \psi)/dr$ are given functions of $r$. 

Next, let us assume that $r_{,u} = 0$ and $r = \alpha$, where $\alpha$, $0 < |\alpha| <1$ is a constant. In this case, Eq.~\eqref{V5} reduces to 
\begin{equation}\label{V6}
(1-\alpha^2) \frac{\phi_{,uu}}{\phi} +  \left(\frac{\psi_{,u}}{\psi}\right)^2 = 0\,.
\end{equation}
We choose a special solution of this equation for further study in Section V; that is,  
\begin{equation}\label{V7}
\phi(u) = \cos (\eta_0 u + \varphi_0)\,,\qquad \psi (u) = e^{\pm \gamma \eta_0 u}\,, \qquad \gamma := (1-\alpha^2)^{\frac{1}{2}}\,,
\end{equation}
where $\eta_0$ and $\varphi_0$ are constants. 
Under the transformations $t \mapsto -t$, $z \mapsto -z$ and $\varphi_0 \mapsto -\varphi_0$,  the special solution with the upper sign turns into the special solution with the lower sign since $u \mapsto -u$.  Henceforth, we work with the solution with the upper sign with no loss in generality.  It is then straightforward to show
\begin{equation}\label{V8}
    \mathbb{K}_1=2\,\eta_0^2 [ \alpha^2 + \gamma\,\tan (\eta_0 u+ \varphi_0)]\,,\qquad \mathbb{K}_2= 2\,\eta_0^2 \alpha [\gamma - \,\tan (\eta_0 u+ \varphi_0)]\,. 
\end{equation}
The energy density of the wave is given by   
\begin{equation}\label{V9}
\rho_g = 8 \ell^2 \eta_0^4  [\alpha^2 +\tan^2 (\eta_0 u+ \varphi_0)]\,. 
\end{equation}
If  $\eta_0 u+ \varphi_0= \pm (n + \tfrac{1}{2}) \pi$, $n = 0, 1, 2, \cdots$,  we encounter a curvature singularity of spacetime since the quantities in Eqs.~\eqref{V8}--\eqref{V9} diverge.  Furthermore, the nonzero components of the acceleration tensor~\eqref{I18} are given by $\Omega = - \alpha \eta_0$ and we can assume the corresponding rotation angle is given by $\Theta = -\alpha (\eta_0 u +  \varphi_0 )$. 

\section{Properties of Nonsingular Solution~\eqref{U5}--\eqref{U8}}

We consider the nonsingular elliptically polarized plane gravitational wave derived in Section III A and given by the metric
\begin{equation}\label{M1}
ds^2 = -dt^2 +dz^2 + e^{\mathbb{S}}\, dx^2 + 2\, \mathbb{S} \, dx \,dy + e^{-\mathbb{S}} \,dy^2\,, \qquad \mathbb{S} = \sin(\eta u + \varphi)\,,
\end{equation}
where we interpret $\eta > 0$ to be typical of the frequency content of this radiation field.  We are interested in the measurements of fiducial geodesic observers that are spatially at rest and carry the parallel propagated tetrad frame $e^{\mu}{}_{\hat \alpha}$; indeed, using Eqs.~\eqref{U6}--\eqref{U8} the resulting measured curvature components can be obtained from
\begin{equation}\label{M2}
\mathcal{K}_1 = -\tfrac{1}{2} \eta^2\, \sin(\eta u + \varphi) \sin\theta\,,   \qquad \mathcal{K}_2 = \tfrac{1}{2} \eta^2\, \sin(\eta u + \varphi) \cos\theta\,,
\end{equation}
where 
\begin{equation}\label{M3}
\theta =  \cos(\eta u + \varphi) + \varphi'\,.
\end{equation}
 To determine the polarization of this plane wave, we consider the rate of change of $\mathcal{K}_2/\mathcal{K}_1$ as in Eqs.~\eqref{I29}--\eqref{I30} and note that the fiducial observer is at a fixed location $(x, y, z)$ in space so that $dt = du$.  Hence, 
\begin{equation}\label{M4}
\frac{d}{du}\left(\frac{\mathcal{K}_2}{\mathcal{K}_1}\right) = -\frac{\eta\,\sin(\eta u + \varphi)}{\sin^2\theta}\,,
\end{equation}
which means that  the radiation for $\eta > 0$ has right-handed polarization when $\sin(\eta u + \varphi) < 0$ and left-handed polarization when $\sin(\eta u + \varphi) > 0$. That is, in the former (latter) case, the GEM fields rotate in the positive (negative) sense  about the direction of wave propagation. 

The frequency content of the measured components of curvature can be determined via Fourier transformation. Inspection of Eqs.~\eqref{M2}--\eqref{M3} reveals that the result in this case is rather complicated. Nevertheless, it is interesting to see how the frequency content changes under a simple rotation of the frame. 

\subsection{Rotating Observer}

Let us therefore imagine that this fiducial observer decides to refer its observations to the tetrad frame $\tilde{e}^{\mu}{}_{\hat \alpha}$ with spatial axes that rotate uniformly with constant angular speed $\tilde{\Omega} > 0$ about the direction of the propagation of the plane wave. Specifically, we assume $\tilde{e}_{\hat 0} = \partial_t$, $\tilde{e}_{\hat 3} = \partial_z$, 
\begin{equation}\label{M5}
\tilde{e}_{\hat 1} = e_{\hat 1}\,\cos (\tilde{\Omega}u)+ e_{\hat 2}\,\sin (\tilde{\Omega}u)\,, \qquad   \tilde{e}_{\hat 2} = - e_{\hat 1}\,\sin (\tilde{\Omega}u)+ e_{\hat 2}\,\cos (\tilde{\Omega}u)\,.
\end{equation}
Under this rotation, we find that the components of the curvature as measured by the observer can now be obtained from 
\begin{equation}\label{M6}
\tilde{\mathcal{K}}_1 = \mathcal{K}_1\,\cos (2\tilde{\Omega}u)+ \mathcal{K}_2\,\sin (2\tilde{\Omega}u)\,, \qquad   \tilde{\mathcal{K}}_2 = - \mathcal{K}_1\,\sin (2\tilde{\Omega}u)+ \mathcal{K}_2\,\cos (2\tilde{\Omega}u)\,.
\end{equation}
Therefore, 
\begin{equation}\label{M7}
\tilde{\mathcal{K}}_1 = -\tfrac{1}{2} \eta^2\, \sin(\eta u + \varphi) \sin \tilde{\theta}\,,    \qquad \tilde{\mathcal{K}}_2 =  \tfrac{1}{2} \eta^2\, \sin(\eta u + \varphi) \cos \tilde{\theta}\,,  
\end{equation}
where $\tilde{\theta} = \theta - 2 \tilde{\Omega} u$. More explicitly,  
\begin{equation}\label{M8}
\tilde{\theta} =  \cos(\eta u + \varphi) + \varphi'- 2 \tilde{\Omega} u\,.
\end{equation}
It follows from the inspection of $\tilde{\mathcal{K}}_1$ and $\tilde{\mathcal{K}}_2$ that the frequency $\eta > 0$ can appear as $\eta - 2 \tilde{\Omega}$ and $\eta + 2 \tilde{\Omega}$, but this is not generally the case.  
 This circumstance should be compared and contrasted with linearized GR. The measured frequency of the wave $\omega$ in the linearized theory of Section I changes to $\tilde{\omega} = \omega \mp 2 \tilde{\Omega}$ for the rotating observer depending upon the helicity of the incident linearized gravitational radiation~\cite{Ramos:2006sb}. Here, the upper (lower) sign indicates incident right (left) circularly polarized plane gravitational wave; moreover, the measured frequency is in accordance with the spin-2 nature of linearized gravitational radiation and the phenomenon of spin-rotation coupling.   Beyond linearized GR, the situation clearly becomes rather complicated and eludes a similar  interpretation.

\subsection{Timelike and Null Geodesics}

We are interested in timelike and null geodesics in this spacetime with 4-velocity vectors  $U^\mu = dx^\mu/d\lambda$, where $\lambda$ is the proper time $\tau$ along a timelike geodesic world line or the affine parameter along a null geodesic. Here,  $U^\mu U_\mu = -\varepsilon$, where $\varepsilon$ is unity (zero) for timelike (null) geodesics. The projection of $U^\mu$ on a Killing vector field is a constant of the motion. For $\partial_x$, $\partial_y$ and $\partial_t + \partial_z$, we have 
\begin{equation}\label{J1}
\partial_x \cdot U =  e^\mathbb{S} \, \frac{dx}{d\lambda} +  \mathbb{S} \, \frac{dy}{d\lambda} = C_1\,,
\end{equation}
\begin{equation}\label{J2}
\partial_y \cdot U = \mathbb{S} \, \frac{dx}{d\lambda} + e^{-\mathbb{S}} \, \frac{dy}{d\lambda} = C_2\,,
\end{equation}
\begin{equation}\label{J3}
(\partial_t + \partial_z) \cdot U = -\frac{dt}{d\lambda} +  \frac{dz}{d\lambda} = - C_u\,, \qquad \frac{du}{d\lambda} = C_u \ne 0\,,
\end{equation}
where $C_1$, $C_2$ and $C_u \ne 0$ are constants. From Eqs.~\eqref{J1} and~\eqref{J2}, we find
\begin{equation}\label{J4}
\frac{dx}{d\lambda} = \frac{C_1 e^{-\mathbb{S}} -C_2 \mathbb{S}}{1-\mathbb{S}^2}\,, \qquad   \frac{dy}{d\lambda} = \frac{-C_1 \mathbb{S} + C_2 e^\mathbb{S}}{1-\mathbb{S}^2}\,.
\end{equation}
Moreover, using these relations in $U^\mu U_\mu = -\varepsilon$, we find after some algebra
\begin{equation}\label{J5}
C_u\,\frac{d(t+z)}{d\lambda} =  \varepsilon + \frac{C^2_1 e^{-\mathbb{S}}+ C^2_2 e^\mathbb{S} - 2 C_1 C_2 \mathbb{S}}{1-\mathbb{S}^2}\,.
\end{equation}
This relation together with $du/d\lambda = C_u \ne 0$ implies
\begin{equation}\label{J6}
\frac{dt}{d\lambda} = \frac{1}{2}\left(\frac{\varepsilon}{C_u} + C_u\right) + \frac{1}{2C_u}\,\frac{C^2_1 e^{-\mathbb{S}}+ C^2_2 e^\mathbb{S} - 2 C_1 C_2 \mathbb{S}}{1-\mathbb{S}^2}\,,
\end{equation}
\begin{equation}\label{J7}
 \frac{dz}{d\lambda} = \frac{1}{2}\left(\frac{\varepsilon}{C_u} - C_u\right) + \frac{1}{2C_u}\,\frac{C^2_1 e^{-\mathbb{S}}+ C^2_2 e^\mathbb{S} - 2 C_1 C_2 \mathbb{S}}{1-\mathbb{S}^2}\,.
\end{equation}

Let us mention the limiting situation that occurs for $C_1 = C_2 = 0$ and $C_u = 1$; in this case, for $\varepsilon = 1$ we recover the 4-velocity vectors of preferred observers that are all spatially at rest in spacetime, while for $\varepsilon =0$ we find the null geodesics that move in the negative $z$ direction opposite to the direction of wave propagation. For the null geodesic that moves along with the wave, we must assume $C_u = C_1 = C_2 = 0$.

Finally, it is interesting to present explicitly the solution of the geodesic equation. To simplify matters, we assume $\varphi = 0$ in Eq.~\eqref{M1} and $u = C_u \lambda$. Moreover, we define $I_{\pm}$ and $I_0$ such that 
\begin{equation}\label{J8}
I_{\pm}(w) = \int_{0}^{w}\frac{e^{\pm \sin \theta}}{\cos^2 \theta}\,d\theta \,, \qquad I_0 (w) =1-\frac{1}{\cos w}\,.
\end{equation}
Then, we can write 
\begin{equation}\label{J11}
t(\lambda) - t(0)  = \frac{\varepsilon+C_u^2}{2\,C_u}\lambda + \frac{1}{2\eta\, C^2_u} [C_1^2\, I_{-}(\eta C_u \lambda) + C_2^2\,I_{+}(\eta C_u \lambda) + 2 C_1C_2\, I_{0}(\eta C_u \lambda)]\,,
\end{equation}
\begin{equation}\label{J12}
x(\lambda) - x(0)  = \frac{C_1}{\eta C_u} I_{-}(\eta C_u \lambda) + \frac{C_2}{\eta C_u} I_{0}(\eta C_u \lambda)\,,
\end{equation}
\begin{equation}\label{J13}
y(\lambda) - y(0)  = \frac{C_2}{\eta C_u} I_{+}(\eta C_u \lambda) + \frac{C_1}{\eta C_u} I_{0}(\eta C_u \lambda)\,,
\end{equation}
\begin{equation}\label{J14}
z(\lambda) - z(0)  = \frac{\varepsilon- C_u^2}{2\,C_u}\lambda + \frac{1}{2\eta\, C^2_u} [C_1^2\, I_{-}(\eta C_u \lambda) + C_2^2\,I_{+}(\eta C_u \lambda) + 2 C_1C_2\, I_{0}(\eta C_u \lambda)]\,.
\end{equation}

\subsection{Cosmic Jet}

A stationary gravitational field is invariant under translation in time and this implies that the energy of a free test particle is conserved; that is, there exists a timelike Killing vector field and the projection of the 4-velocity of the particle on the Killing vector is in essence the particle's energy and is a constant of the motion.  In a time-varying gravitational field, however, free test particles can gain or lose energy along their world lines. Two aspects of this energy exchange with the gravitational field have been previously investigated that result in cosmic jets. The first scenario involves certain dynamic spacetime domains where collapse occurs along some direction; then,  free test particles are accelerated along the direction of collapse relative to comoving observers that are all spatially at rest in spacetime. In such time-dependent gravitational fields, parallel and antiparallel to the collapsing direction, double-jet outflows appear in which the speeds of free test particles can asymptotically approach the  speed of light~\cite{Chicone:2010aa, Chicone:2010hy, Chicone:2010xr}. The second scenario involves certain gravitational wave spacetimes where  most of the free test particles asymptotically line up relative to the comoving observers and form a single-jet structure such that the speed of the jet asymptotically approaches the speed of light~\cite{Bini:2014esa, Bini:2017qnd}. These cosmic-jet scenarios are theoretical constructs that appear to be distinct from astrophysical jets. 

The cosmic-jet phenomenon has been studied in exact \emph{linearly polarized} plane gravitational wave spacetimes, where the jet motion is in the same direction as the plane wave~\cite{Bini:2014esa, Bini:2017qnd}. Moreover, this circumstance occurs as well in twisted nonplanar unidirectional gravitational wave spacetimes, but the direction of jet motion is different from the direction of wave propagation~\cite{Bini:2018gbq, Firouzjahi:2019qmy}.  We now wish to extend the study of cosmic jets to the case of exact \emph{elliptically polarized} gravitational plane waves.

For  $\varepsilon = 1$, $C_u = 1$ and $C_1= C_2 = 0$, we have the preferred geodesic observers that are all fixed in space.  These preferred observers have adapted orthonormal tetrads $\chi^{\mu}{}_{\hat \alpha}$ given in Eq.~\eqref{I7}. In the present case, we can express the tetrad components in $(t, x, y, z)$ coordinates as
\begin{equation}\label{J15}
\chi^{\mu}{}_{\hat 0} =  (1, 0, 0, 0)\,, \qquad \chi^{\mu}{}_{\hat 1} = (0, e^{-\tfrac{1}{2}\, \mathbb{S}}, 0, 0)\,,
\end{equation}
\begin{equation}\label{J16}
\chi^{\mu}{}_{\hat 2} = (1- \mathbb{S}^2)^{-1/2}\,(0, - \mathbb{S}\,e^{-\tfrac{1}{2}\, \mathbb{S}}, e^{\tfrac{1}{2}\, \mathbb{S}}, 0)\,,\qquad \chi^{\mu}{}_{\hat 3} = (0, 0, 0, 1)\,,
\end{equation}
where $\mathbb{S}^2 < 1$ by assumption; indeed, the coordinate system becomes inadmissible for $ \mathbb{S}^2 = 1$. According to the preferred observers, the 4-velocity vector of an arbitrary timelike or null geodesic is given by
\begin{equation}\label{J17}
U_{\hat \alpha} =  U_\mu \, \chi^{\mu}{}_{\hat \alpha}\,,
\end{equation}
where $U^\mu = dx^\mu /d\lambda$ and either $\lambda$ is the proper time $\tau$ of the timelike geodesic motion with $\varepsilon = 1$ or $\lambda$ is the affine parameter along a null geodesic with $\varepsilon = 0$. In either case, we find explicitly
\begin{equation}\label{J18}
U^{\hat \alpha} =  \left(\frac{dt}{d\lambda}, C_1 e^{-\tfrac{1}{2}\, \mathbb{S}}, (1- \mathbb{S}^2)^{-1/2}\,e^{-\tfrac{1}{2}\, \mathbb{S}}\,(-C_1 \mathbb{S} + C_2e^ \mathbb{S}), \frac{dz}{d\lambda}\right)\,, 
\end{equation}
where $dt/d\lambda$ and $dz/d\lambda$ are given by Eqs.~\eqref{J6} and~\eqref{J7}, respectively.  The 4-velocity of a timelike or null geodesic world line as measured by the reference observers can be expressed in terms of the measured velocity components $(V_{\hat 1}, V_{\hat 2}, V_{\hat 3})$ in the form
\begin{equation}\label{J19}
U^{\hat \alpha} =  \Gamma ( 1, V_{\hat 1}, V_{\hat 2}, V_{\hat 3})\,, 
\end{equation}
where $\Gamma = dt/d\tau$ is the Lorentz factor in the case of a timelike world line. It follows from Eqs.~\eqref{J6} and~\eqref{J7} that  as $\mathbb{S}^2 \to 1$, a cosmic jet develops for arbitrary $C_1$ and $C_2$  involving timelike and null geodesics that line up along the direction of wave propagation such that $\Gamma \to \infty$ and
\begin{equation}\label{J20}
(V_{\hat 1}, V_{\hat 2}, V_{\hat 3}) \to ( 0, 0, 1)\,. 
\end{equation}
The jet structure has been invariantly specified; therefore, it is physically meaningful. 

The cosmic jet develops relative to the preferred geodesic observers that are all spatially at rest and therefore do not participate in the jet formation process. Indeed,  some other geodesics may also escape the fate of the majority for specific choices of the parameters; for instance, in the case under consideration, if $C_1 = e \,C_2$ ( $C_2 = - e\, C_1$), all of the components of $U^{\hat \alpha}$ remain finite as $\mathbb{S} \to 1$ ($\mathbb{S} \to -1$).

\subsection{Geodesic Motion in the Fermi Frame}

The geodesic equations of motion~\eqref{G9}--\eqref{G11} in the Fermi frame are in general nonlinear. If we drop the velocity terms, the generalized Jacobi equation reduces to the Jacobi equation that is linear. The resulting geodesic deviation equations in the present case are then given by
\begin{equation}\label{K1}
 \frac{d^2X}{du^2} + \mathcal{K}_1 X + \mathcal{K}_2 Y = 0\,, 
\end{equation}
\begin{equation}\label{K2}
 \frac{d^2Y}{du^2} + \mathcal{K}_2 X - \mathcal{K}_1 Y = 0\, 
\end{equation}
and $d^2Z/du^2 = 0$, since $dt = du$ for the fiducial observer. Let us assume $Z = 0$ and focus on motion in the $(X, Y)$ plane with $\mathcal{K}_1$ and $\mathcal{K}_2$ given by Eqs.~\eqref{M2}--\eqref{M3}. At the spatial origin of Fermi coordinates, the speeds of the free test particles or null rays must be less than or equal to unity, respectively. In this connection, let us define $(\mathcal{X}, \mathcal{Y}) = (\eta X, \eta Y)$ and note that $(dX/du, dY/du) = (d\mathcal{X}/d\varpi,   d\mathcal{Y}/d\varpi)$, where 
$\varpi := \eta u + \varphi$. In Eq.~\eqref{M3}, let $\varphi' = 0$ for simplicity and note that Eqs.~\eqref{K1} and~\eqref{K2} can then be written as
\begin{equation}\label{K3}
 \frac{d^2\mathcal{X}}{d\varpi^2} -\tfrac{1}{2}\,\sin \varpi\, \sin(\cos \varpi ) \mathcal{X} +\tfrac{1}{2}\,\sin \varpi\, \cos(\cos \varpi ) \mathcal{Y} = 0\,, 
\end{equation}
\begin{equation}\label{K4}
 \frac{d^2\mathcal{Y}}{d\varpi^2} + \tfrac{1}{2}\,\sin \varpi\, \cos(\cos \varpi ) \mathcal{X} +\tfrac{1}{2}\,\sin \varpi\, \sin(\cos \varpi ) \mathcal{Y} = 0\,.  
\end{equation}
To illustrate the behavior of this system, we numerically integrate these equations starting from $\varpi = 0$ with specific initial conditions for $(\mathcal{X}, \mathcal{Y})$ and $(d\mathcal{X}/d\varpi,   d\mathcal{Y}/d\varpi)$ as in Figure 1.  The coordinate system is admissible up to $\varpi = \pi/2$; however, the integration can continue smoothly beyond this point, since no impediment is encountered in the invariantly defined Jacobi equation. Figure 1 represents the resulting motion in the $(\mathcal{X}, \mathcal{Y})$ plane for two possible sets of initial conditions. 

Let us next consider free test particles that are initially at rest on a circle about the origin of spatial Fermi coordinates in the $(X, Y)$ plane. The motion of the particles as a consequence of the passage of the plane gravitational wave can be calculated in principle based on the linear system~\eqref{K1}--\eqref{K2}. The initial circular pattern in time becomes a time-varying  ellipse, since circles go to ellipses under affine maps on finite dimensional spaces. 

\begin{figure}[h]
\subfloat[]{\includegraphics[width = 3in]{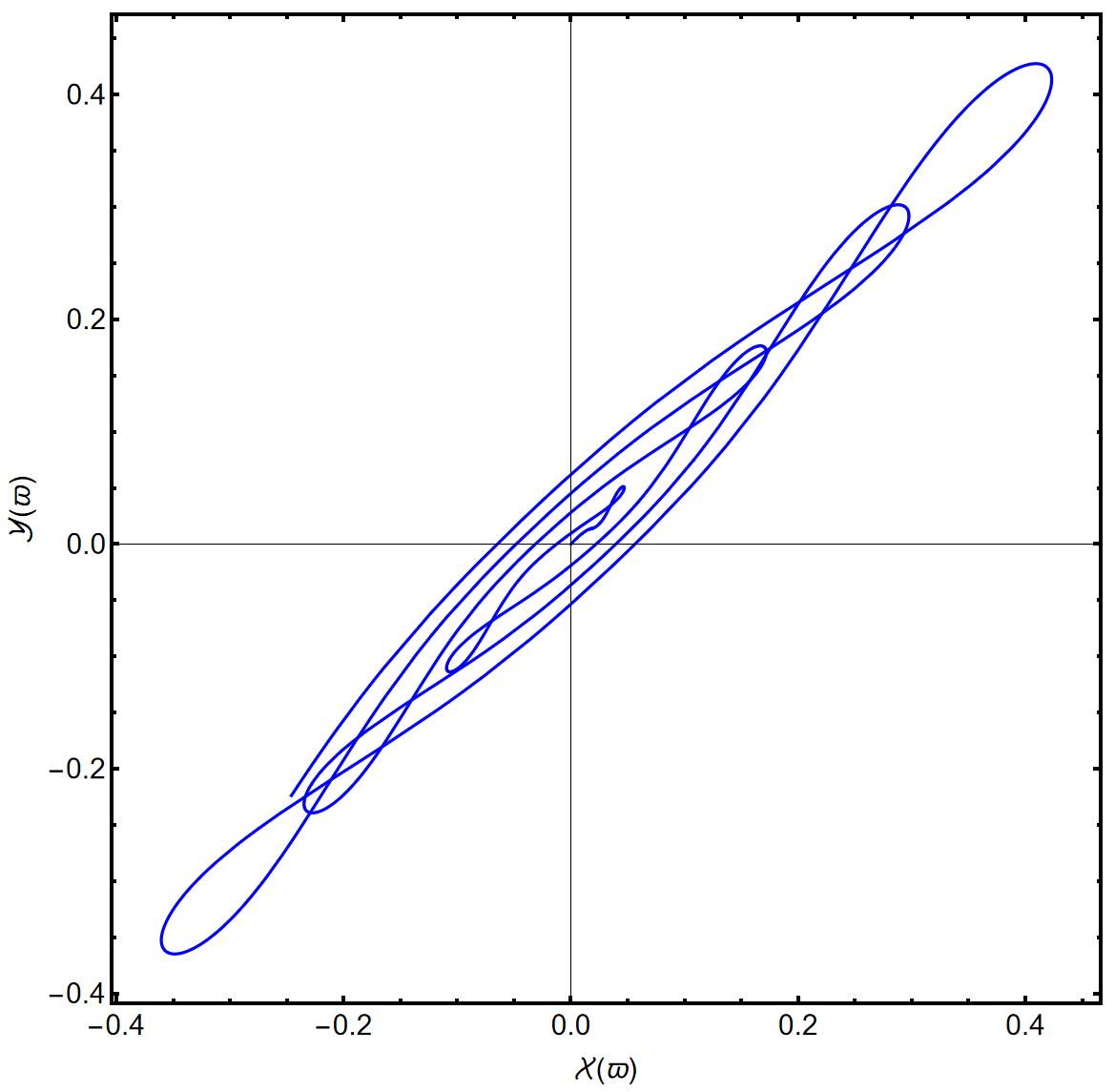}} 
\subfloat[]{\includegraphics[width = 3in]{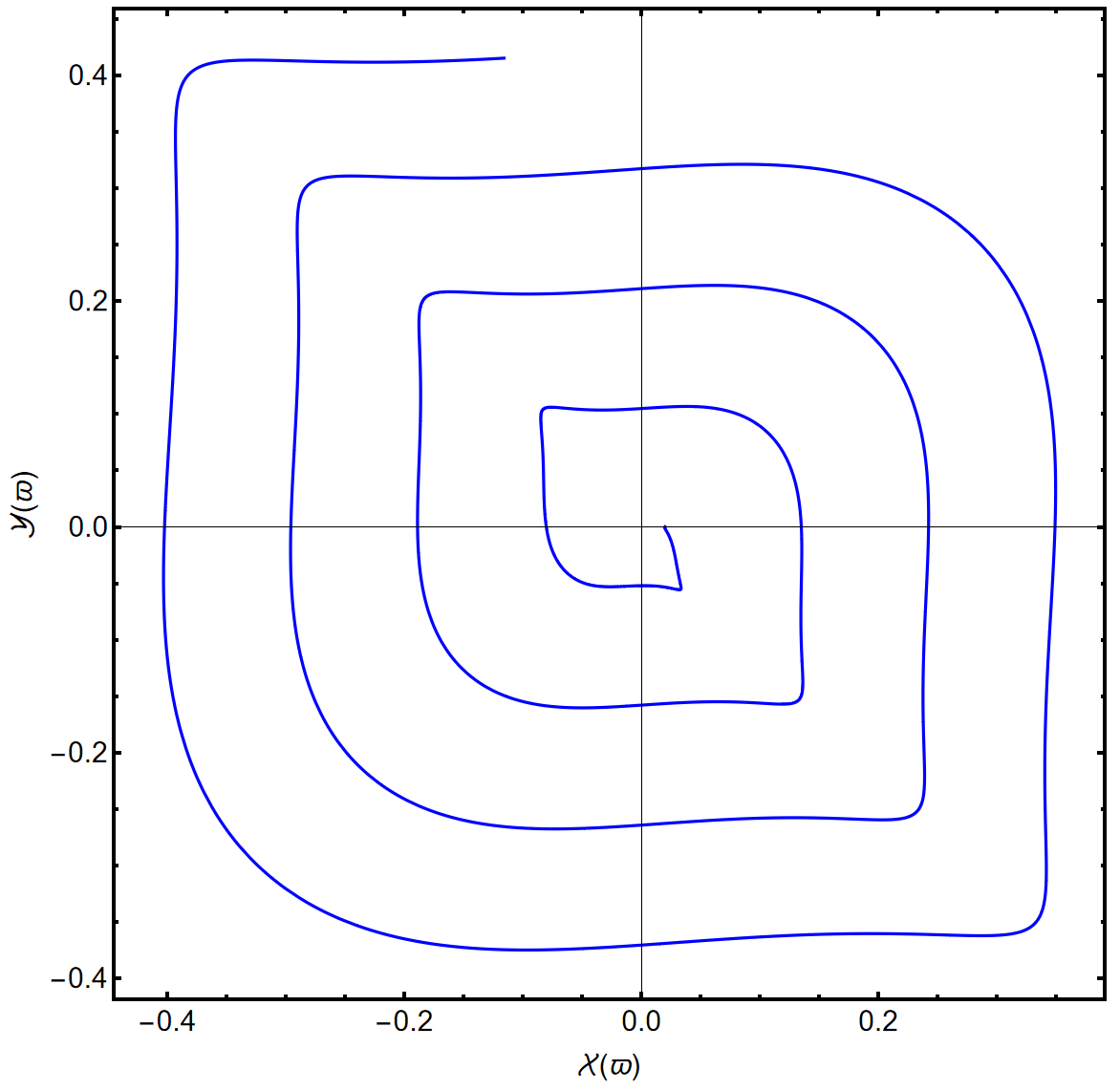}}
\caption{The results of the numerical integration of system~\eqref{K3}--\eqref{K4} for $0<\varpi<50$ with initial conditions (a): $(\mathcal{X}, \mathcal{Y})= (0, 0)$, $(d\mathcal{X}/d\varpi, d\mathcal{Y}/d\varpi) = (0.01, 0.01)$ and (b): $(\mathcal{X},\mathcal{Y}) = (0.02, 0)$, $(d\mathcal{X}/d\varpi, d\mathcal{Y}/d\varpi) = (-10^{-3}, -10^{-3}).$}
\label{fig:1}
\end{figure}

\section{Properties of Singular Solution~\eqref{V7}--\eqref{V9}}

Let us next consider the particular solution given by the metric
\begin{equation}\label{N1}
ds^2 = - dt^2 + dz^2 + \cos^2(\eta_0 u +  \varphi_0 )\,\left(e^{2 \gamma \eta_0  u} \,dx^2 + 2 \alpha\,dx dy + e^{- 2 \gamma \eta_0 u} \,dy^2\right)\,,
\end{equation}
where $\eta_0$, $\varphi_0$ and $\alpha$ are constants. Here, $\gamma = (1-\alpha^2)^{\frac{1}{2}}>0$. This spacetime has a curvature  singularity where $\cos(\eta_0 u +  \varphi_0 ) = 0$; moreover, the coordinate system is admissible for
$0 < \alpha^2 < 1$.   As in the previous section,  the wave characteristics can be studied in this somewhat more complicated situation; however, for the sake of simplicity, we limit our considerations here to the timelike and null geodesics and the formation of the cosmic jet. Let   $\mathbb{U}^\mu = dx^\mu/d\lambda_0$ be the 4-velocity vector of a geodesic world line, where $\lambda_0$ is the proper time $\tau_0$ of a timelike geodesic or  the affine parameter of a null geodesic; as before, $\mathbb{U}^\mu \mathbb{U}_\mu = -\varepsilon$, where $\varepsilon$ is unity  or zero, respectively. We project 
$\mathbb{U}^\mu$ on the  Killing vector fields $\partial_x$, $\partial_y$ and $\partial_t + \partial_z$ to find the equations of motion. Specifically,  
\begin{equation}\label{N2}
\partial_x \cdot \mathbb{U} =  \cos^2(\eta_0 u +  \varphi_0 )\,\left(e^{2 \gamma \eta_0 u} \, \frac{dx}{d\lambda_0} +  \alpha \, \frac{dy}{d\lambda_0}\right) = D_1\,,
\end{equation}
\begin{equation}\label{N3}
\partial_y \cdot \mathbb{U} =  \cos^2(\eta_0 u +  \varphi_0 )\,\left(\alpha \, \frac{dx}{d\lambda_0} + e^{-2 \gamma \eta_0 u} \, \frac{dy}{d\lambda_0}\right) = D_2\,,
\end{equation}
\begin{equation}\label{N4}
(\partial_t + \partial_z) \cdot \mathbb{U} = -\frac{dt}{d\lambda_0} +  \frac{dz}{d\lambda_0} = - D_u\,, \qquad \frac{du}{d\lambda_0} = D_u \ne 0\,,
\end{equation}
where $D_1$, $D_2$ and $D_u \ne 0$ are constants of the motion. Eqs.~\eqref{N2} and~\eqref{N3} imply
\begin{equation}\label{N5}
\frac{dx}{d\lambda_0} = \frac{e^{-2 \gamma \eta_0 u}D_1  - \alpha D_2 }{\gamma^2\,\cos^2(\eta_0 u +  \varphi_0 )}\,, \qquad   \frac{dy}{d\lambda_0} = \frac{- \alpha D_1 +  e^{2 \gamma \eta_0 u}D_2}{\gamma^2\,\cos^2(\eta_0 u +  \varphi_0 )}\,.
\end{equation}
Next, from $\mathbb{U}^\mu \mathbb{U}_\mu = -\varepsilon$, we find
\begin{equation}\label{N6}
D_u\,\frac{d(t+z)}{d\lambda_0} =  \varepsilon + \frac{e^{-2 \gamma \eta_0 u}D^2_1 + e^{2 \gamma \eta_0 u}D^2_2  - 2 \alpha D_1 D_2}{\gamma^2\, \cos^2(\eta_0 u +  \varphi_0 )}\,.
\end{equation}
Employing $du/d\lambda_0 = D_u \ne 0$, we now have 
\begin{equation}\label{N7}
\frac{dt}{d\lambda_0} = \frac{1}{2}\left(\frac{\varepsilon}{D_u} + D_u\right) + \frac{1}{2D_u}\,\frac{e^{-2 \gamma \eta_0 u}D^2_1 + e^{2 \gamma \eta_0 u}D^2_2  - 2 \alpha D_1 D_2}{\gamma^2\, \cos^2(\eta_0 u +  \varphi_0 )}\,,
\end{equation}
\begin{equation}\label{N8}
 \frac{dz}{d\lambda_0} = \frac{1}{2}\left(\frac{\varepsilon}{D_u} - D_u\right) + \frac{1}{2D_u}\,\frac{e^{-2 \gamma \eta_0 u}D^2_1 + e^{2 \gamma \eta_0 u}D^2_2  - 2 \alpha D_1 D_2}{\gamma^2\, \cos^2(\eta_0 u +  \varphi_0 )}\,.
\end{equation}

We are interested in the motion of these spacetime geodesics as determined by preferred observers that are all spatially at rest in this spacetime. The fiducial observers indeed have timelike geodesic world lines with 4-velocity vectors given by $D_1 = D_2 = 0$, $D_u = 1$ and $\varepsilon = 1$. In this case, the natural tetrad frame adapted to the fiducial observers, namely, $\chi^{\mu}{}_{\hat \alpha}$ given in Eq.~\eqref{I7} can be expressed  in $(t, x, y, z)$ coordinates as
\begin{equation}\label{N9}
\chi^{\mu}{}_{\hat 0} =  (1, 0, 0, 0)\,, \qquad \chi^{\mu}{}_{\hat 1} = (0, \frac{e^{-\gamma \eta_0 u}}{\cos(\eta_0 u +  \varphi_0 )}, 0, 0)\,,
\end{equation}
\begin{equation}\label{N10}
\chi^{\mu}{}_{\hat 2} = \frac{1}{\gamma\,\cos(\eta_0 u +  \varphi_0 )}\,(0, - \alpha\,e^{-\gamma \eta_0 u}, e^{\gamma \eta_0 u}, 0)\,,\qquad \chi^{\mu}{}_{\hat 3} = (0, 0, 0, 1)\,.
\end{equation}
The 4-velocity vector of an arbitrary timelike or null geodesic as determined by the fiducial observers is given by
\begin{equation}\label{N11}
\mathbb{U}_{\hat \alpha} =  \mathbb{U}_\mu \, \chi^{\mu}{}_{\hat \alpha}\,.
\end{equation}
Using Eqs.~\eqref{N5}--\eqref{N10}, we find
\begin{equation}\label{N12}
\mathbb{U}^{\hat \alpha} =  \left(\frac{dt}{d\lambda_0},\, \frac{D_1 e^{-\gamma \eta_0 u}}{\cos(\eta_0 u +  \varphi_0 )},\, \frac{- \alpha\,D_1e^{-\gamma \eta_0 u}+D_2 e^{\gamma \eta_0 u}}{\gamma\,\cos(\eta_0 u +  \varphi_0 )},\, \frac{dz}{d\lambda_0}\right)\,, 
\end{equation}
where $dt/d\lambda_0$ and $dz/d\lambda_0$ are given by Eqs.~\eqref{N7} and~\eqref{N8}, respectively.   The 4-velocity vector as measured by the reference observers can be expressed in terms of the measured velocity components 
$(\mathbb{V}_{\hat 1}, \mathbb{V}_{\hat 2}, \mathbb{V}_{\hat 3})$ in the form
\begin{equation}\label{N13}
\mathbb{U}^{\hat \alpha} =  \Gamma_0 (1, \mathbb{V}_{\hat 1}, \mathbb{V}_{\hat 2}, \mathbb{V}_{\hat 3})\,, 
\end{equation}
where $\Gamma_0 = dt/d\tau_0$ is the Lorentz factor in the case of a timelike world line.  Eqs.~\eqref{N7}--\eqref{N8} and~\eqref{N12} imply that for arbitrary $D_1$ and $D_2$ a cosmic jet develops involving timelike and null geodesics that line up along the direction of wave propagation as the singularity is approached for $\cos(\eta_0 u +  \varphi_0 ) \to 0$; that is, in this limit $\Gamma_0 \to \infty$ and
\begin{equation}\label{N14}
(\mathbb{V}_{\hat 1}, \mathbb{V}_{\hat 2}, \mathbb{V}_{\hat 3}) \to ( 0, 0, 1)\,.
\end{equation}


\section{Discussion}

We have studied the physics of nonlinear elliptically polarized plane gravitational waves within the framework of GR. The wave properties can be described within the GEM framework in analogy with electromagnetic waves; moreover, we deal with the energetics of gravitational plane waves via the GEM energy-momentum tensor. Exact Ricci-flat solutions of GR representing gravitational plane waves are displayed and two specific elliptically polarized examples are chosen for further analysis. We solve the geodesic equations of motion in these cases and demonstrate the cosmic jet property. That is, as the waves propagate, most of the timelike (as well as null) geodesics line up, as measured by fiducial observers that are spatially at rest in spacetime, and thereby produce a cosmic jet whose speed asymptotically approaches the speed of light. 

Nonlinear gravitational waves in GR carry energy, momentum and angular momentum. Our brief GEM treatment amounts to replacing the plane wave with a bundle of parallel null rays with a certain energy density. Alternatively, one could deal with these concepts within the framework of the teleparallel equivalent of GR~\cite{Obukhov:2009gv, Maluf:2023rwe, Maluf:2024bwg, Costa:2025jkw}. A detailed consideration of these aspects of exact plane gravitational waves is beyond the scope of the present work and remains a task for the future.  

\section*{ACKNOWLEDGMENTS}
SB is partially supported by the Abdus Salam International Center for Theoretical Physics (ICTP) under the regular associateship scheme.
Moreover, MM and SB are partially supported by the Sharif University of Technology Office of Vice President for Research under Grant No. G4010204. 


\appendix

\section{GEM Energy-Momentum Tensor}

Within the GEM framework, it is possible to define a certain energy-momentum tensor in close analogy with electrodynamics. This invariant average stress-energy tensor of a gravitational field is constructed using the Riemannian curvature of spacetime via the Bel superenergy tensor~\cite{Mashhoon:1996wa, Mashhoon:1998tt, Mashhoon:2000he, Mashhoon:2003ax, Bini:2018lhh}. In a Ricci-flat gravitational field, the Riemann tensor reduces to the Weyl conformal
tensor $C_{\mu\nu\rho \sigma}$ and the Bel tensor reduces to the completely symmetric and traceless Bel-Robinson tensor $T_{\mu\nu\rho \sigma}$ given by
\begin{equation}\label{B1} 
T_{\mu\nu\rho \sigma} = \frac{1}{2} (C_{\mu\xi\rho \zeta} C_{\nu\;\; \sigma}^{\;\; \xi \;\;\zeta}+C_{\mu\xi\sigma \zeta}C_{\nu\;\;\rho}^{\;\;\xi \;\; \zeta})-\frac{1}{16}g_{\mu\nu} g_{\rho \sigma}C_{\alpha \beta\gamma \delta}C^{\alpha \beta\gamma \delta}\,.
\end{equation}
We define the GEM energy-momentum tensor $\mathcal{T}_{\hat \alpha \hat \beta}$ of a Ricci-flat gravitational field by
\begin{equation}\label{B2} 
\mathcal{T}_{\hat \alpha \hat \beta}=\ell^2\, T_{\mu\nu\rho \sigma }\,e^{\mu}{}_{\hat \alpha}\,e^{\nu}{}_{\hat \beta}\,e^{\rho}{}_{\hat 0}\, e^{\sigma}{}_{\hat 0},
\end{equation}
where $\ell$ is a constant intrinsic  lengthscale associated with the gravitational field and $e^{\mu}{}_{\hat \alpha}$ is a parallel-propagated tetrad frame adapted to a free fiducial observer following a timelike geodesic in spacetime. As in electrodynamics,  the GEM energy-momentum tensor is traceless. 

Expressing the gravitoelectric and gravitomagnetic components of the Weyl curvature tensor by $\mathcal{E}$ and $\mathcal{B}$, respectively, as in Eq.~\eqref{6}, we find
\begin{equation}\label{B3} 
\mathcal{T}_{\hat 0 \hat 0} = \tfrac{1}{2}\,\ell^2\, \mathrm{tr}(\mathcal{E}^2 + \mathcal{B}^2)\,, \qquad \mathcal{T}_{\hat 0 \hat i} = \ell^2\,\epsilon_{ijk} (\mathcal{E}\,\mathcal{B})_{jk}\,, 
\end{equation}
\begin{equation}\label{B4} 
\mathcal{T}_{\hat i \hat j} = \ell^2\, [ -\,(\mathcal{E}^2 + \mathcal{B}^2)_{ij} + \tfrac{1}{2}\,\delta_{ij}\, \mathrm{tr}(\mathcal{E}^2 + \mathcal{B}^2)]\,.
\end{equation}

For the general plane gravitational waves under consideration in this paper,  
\begin{equation}
\label{B5}
\mathcal{E}= \left[
\begin{array}{ccc}
\mathcal{K}_1&\mathcal{K}_2 & 0\cr
\mathcal{K}_2&-\mathcal{K}_1 & 0\cr
0&0 & 0\cr
\end{array}
\right]\,,\qquad
\mathcal{B}= \left[
\begin{array}{ccc}
\mathcal{K}_2&-\mathcal{K}_1 & 0\cr
-\mathcal{K}_1&-\mathcal{K}_2 & 0\cr
0&0 & 0\cr
\end{array}
\right]\,.
\end{equation}
Employing these expressions in Eqs.~\eqref{B3}--\eqref{B4}, we find
\begin{equation}\label{B6} 
\mathcal{T}^{\hat 0 \hat 0} = \rho_g\,, \qquad  \mathcal{T}^{\hat 0 \hat i} = \rho_g\,\delta^i_3\,, \qquad \mathcal{T}^{\hat i \hat j} = \rho_g\,\delta^i_3\,\delta^j_3\,,
\end{equation}
where $\rho_g$,  the energy density of the gravitational plane wave in the GEM framework,  is given by
\begin{equation}\label{B7} 
\rho_g  : = 2\ell^2\,(\mathcal{K}_1^2 + \mathcal{K}_2^2)\,.
\end{equation}
 Therefore, we can write
\begin{equation}\label{B8} 
\mathcal{T}^{\hat \alpha \hat \beta} = \rho_g\,k^{\hat \alpha}\,k^{\hat \beta}\,, \qquad  k = \partial_t + \partial_z\,,
\end{equation}
where $k$ is the covariantly constant null vector field that represents the propagation vector of the gravitational plane wave in the $z$ direction at the speed of light.

\section{Fermi Connection Coefficients}

In this Appendix, Christoffel symbols $\Gamma^{\hat \mu}_{\hat \nu \hat \rho}$ for the Fermi coordinate system based on metric~\eqref{I2}  will be denoted for simplicity by $\widehat{\Gamma}^{\mu}_{\nu \rho}$. In accordance with our approximation scheme, we ignore Christoffel symbols that are of second or higher order in the spatial Fermi coordinates. The nonzero components of $\widehat{\Gamma}^\mu_{\nu \rho} = \widehat{\Gamma}^\mu_{\rho \nu}$ can be obtained from
\begin{equation}\label{C1}
\widehat{\Gamma}^{0}_{0 1} = - 3\, \widehat{\Gamma}^{0}_{1 3} = - \widehat{\Gamma}^{1}_{0 3}= \widehat{\Gamma}^{3}_{0 1}= \tfrac{3}{2}\, \widehat{\Gamma}^{1}_{3 3}= -3\, \widehat{\Gamma}^{3}_{1 3} = \mathcal{K}_1 X + \mathcal{K}_2 Y\,,
\end{equation}
\begin{equation}\label{C2}
\widehat{\Gamma}^{0}_{0 2} = - 3\, \widehat{\Gamma}^{0}_{2 3} = - \widehat{\Gamma}^{2}_{0 3}= \widehat{\Gamma}^{3}_{0 2}= \tfrac{3}{2}\, \widehat{\Gamma}^{2}_{3 3}= -3\, \widehat{\Gamma}^{3}_{2 3} = \mathcal{K}_2 X - \mathcal{K}_1 Y\,,
\end{equation}
\begin{equation}\label{C3}
\widehat{\Gamma}^{0}_{1 1} = - 2\, \widehat{\Gamma}^{1}_{1 3} = 2\, \widehat{\Gamma}^{2}_{2 3}= -  \widehat{\Gamma}^{0}_{2 2} =  \widehat{\Gamma}^{3}_{1 1}=  - \widehat{\Gamma}^{3}_{2 2} = \tfrac{2}{3}\, \mathcal{K}_1 Z\,,
\end{equation}
\begin{equation}\label{C4}
\widehat{\Gamma}^{0}_{1 2} = - 2\, \widehat{\Gamma}^{1}_{2 3} = - 2\, \widehat{\Gamma}^{2}_{1 3} =  \widehat{\Gamma}^{3}_{1 2} = \tfrac{2}{3}\, \mathcal{K}_2 Z\,.
\end{equation}

 \end{document}